\documentclass[%
 reprint,
superscriptaddress,
 amsmath,amssymb,
 aps,
prb,
floatfix,
]{revtex4-2}

\usepackage{makecell}
\usepackage{graphicx}
\usepackage{dcolumn}
\usepackage{siunitx} 
\usepackage{braket} 
\usepackage{makecell}
\usepackage{ulem} 
\usepackage{xcolor} 
\usepackage[colorlinks=true,citecolor=blue,linkcolor=blue,pdfhighlight =/O]{hyperref}

\newcommand{\eref}[1]{Eq.~(\ref{#1})} 
\newcommand{\sfref}[1]{Figure~\ref{#1}} 
\newcommand{\sfrefs}[1]{Figures~\ref{#1}} 
\newcommand{\fref}[1]{Fig.~\ref{#1}} 
\newcommand{\frefs}[1]{Figs.~\ref{#1}} 

\begin{document}

\preprint{APL Quantum}

\title{Characterizing and Mitigating Flux Crosstalk in Superconducting Qubits-Couplers System}

\author{Myrron Albert Callera Aguila}
\thanks{M. A. C. A., N.-Y. L. and C.-H. M. contributed equally in the manuscript}
\affiliation{Research Center for Critical Issues, Academia Sinica, Guiren, Tainan, 711010, Taiwan}

\author{Nien-Yu Li}
\thanks{M. A. C. A., N.-Y. L. and C.-H. M. contributed equally in the manuscript}
\affiliation{Institute of Physics, Academia Sinica, Nankang, Taipei, 11529, Taiwan}
\affiliation{Department of Physics, National Taiwan University, Da'an District, Taipei 10617, Taiwan}

\author{Chen-Hsun Ma}
\thanks{M. A. C. A., N.-Y. L. and C.-H. M. contributed equally in the manuscript}
\affiliation{Institute of Physics, Academia Sinica, Nankang, Taipei, 11529, Taiwan}
\affiliation{Department of Physics, National Taiwan University, Da'an District, Taipei 10617, Taiwan}

\author{Li-Chieh Hsiao}
\affiliation{Research Center for Critical Issues, Academia Sinica, Guiren, Tainan, 711010, Taiwan}

\author{Yi-Shiang Huang}
\affiliation{Institute of Physics, Academia Sinica, Nankang, Taipei, 11529, Taiwan}

\author{Yen-Chun Chen}
\affiliation{Research Center for Critical Issues, Academia Sinica, Guiren, Tainan, 711010, Taiwan}

\author{Teik-Hui Lee}
\affiliation{Research Center for Critical Issues, Academia Sinica, Guiren, Tainan, 711010, Taiwan}

\author{Chin-Chia Chang}
\affiliation{Institute of Physics, Academia Sinica, Nankang, Taipei, 11529, Taiwan}

\author{Jyh-Yang Wang}
\affiliation{Department of Electrical Engineering, Feng Chia University, Xitun, Taichung 407301, Taiwan}

\author{Ssu-Yen Huang}
\affiliation{Department of Physics, National Taiwan University, Da'an District, Taipei 10617, Taiwan}

\author{Hsi-Sheng Goan}
\affiliation{Department of Physics, National Taiwan University, Da'an District, Taipei 10617, Taiwan}
\affiliation{Center for Quantum Science and Engineering, National Taiwan University, Taipei 106319, Taiwan}
\affiliation{Physics Division, National Center for Theoretical Sciences, Taipei 106319, Taiwan}

\author{Chiao-Hsuan Wang}
\affiliation{Department of Physics, National Taiwan University, Da'an District, Taipei 10617, Taiwan}
\affiliation{Center for Quantum Science and Engineering, National Taiwan University, Taipei 106319, Taiwan}
\affiliation{Physics Division, National Center for Theoretical Sciences, Taipei 106319, Taiwan}

\author{Cen-Shawn Wu}
\affiliation{Department of Physics, National Changhua University of Education, Changhua, Changhua 500207, Taiwan}
\affiliation{Research Center for Critical Issues, Academia Sinica, Guiren, Tainan, 711010, Taiwan}

\author{Chii-Dong Chen}
\affiliation{Institute of Physics, Academia Sinica, Nankang, Taipei, 11529, Taiwan}
\affiliation{Research Center for Critical Issues, Academia Sinica, Guiren, Tainan, 711010, Taiwan}

\author{Chung-Ting Ke}
\email{ctke@as.edu.tw}
\affiliation{Institute of Physics, Academia Sinica, Nankang, Taipei, 11529, Taiwan}
\affiliation{Research Center for Critical Issues, Academia Sinica, Guiren, Tainan, 711010, Taiwan}

\date{\today}

\begin{abstract}

Superconducting qubits have achieved exceptional gate fidelities, exceeding the error-correction threshold in recent years. One key ingredient of such improvement is the introduction of tunable couplers to control the qubit-to-qubit coupling through frequency tuning. Moving toward fault-tolerant quantum computation, increasing the number of physical qubits is another step toward effective error correction codes. Under a multiqubit architecture, flux control (Z) lines are crucial in tuning the frequency of the qubits and couplers. However, dense flux lines result in magnetic flux crosstalk, wherein magnetic flux applied to one element inadvertently affects neighboring qubits or couplers. This crosstalk obscures the idle frequency of the qubit when flux bias is applied, which degrades gate performance and calibration accuracy. In this study, we characterize flux crosstalk and suppress it in a multiqubit-coupler chip with multi-Z lines without adding additional readout for couplers. By quantifying the mutual flux-induced frequency shifts of qubits and couplers, we construct a cancellation matrix that enables precise compensation of non-local flux, demonstrating a substantial reduction in Z-line crosstalk from 56.5$\,$\textperthousand$\,$to 0.13$\,$\textperthousand$\,$ which is close to statistical error. Flux compensation corrects the CZ SWAP measurement, leading to a symmetric map with respect to flux bias. Compared with a crosstalk-free calculated CZ SWAP map, the measured map indicates that our approach provides a near-zero crosstalk for the coupler-transmon system. These results highlight the effectiveness of our approach in enhancing flux crosstalk-free control and supporting its potential for scaling superconducting quantum processors.  

\end{abstract}

\maketitle
Superconducting qubits are one of the promising types of physical qubits to realize fault-tolerant quantum computing (FTQC), in which quantum computers tackle problems that conventional computers cannot solve on a reasonable time scale~\cite{Arute_2019, Mohseni_2025}. In operating high-fidelity quantum gates, the tunable coupler plays a crucial role. It suppresses spurious coupling and modulates the coupling strength between two qubits~\cite{YChen2014, Yan2018, Stehlik_2021} via coupler-mediated virtual exchange~\cite{Yan2018, Arute_2019}. These features enhance single-qubit (1Q) and two-qubit (2Q) gate fidelities in both transmon and fluxonium architectures ~\cite{Sung_2021, Ding_2023} in a superconducting quantum processor.  Reaching the error correction threshold marks the first step towards FTQC ~\cite{Arute_2019, Lacroix_2024, Google_2025, Gao_2025}.  

To achieve FTQC, an increasing number of qubits and couplers is needed, which leads to more control lines as well as crosstalk between qubits and couplers. Crosstalks could originate from unintended microwave~\cite{RWang2022, HYan2023, Yang_2024, XYYang2024} and readout drives~\cite{Heinsoo_2018,Pitsun_2020}, residual qubit-to-qubit coupling~\cite{YChen2014, Yan2018, Sung_2021, RLi2024}, frequency crowding~\cite{Hertzberg2021, Stehlik_2021, EZhang2022, Pappas2024}, and unwanted flux~\cite{Barends2014, Neill2018, Barrett2023, Dai2024} that leads to imperfection of the 1Q and 2Q gates. Flux crosstalk plays a role in a frequency-tunable qubit-coupler system; it obscures the precise frequency detuning required for a quantum gate operation. Precise control of the flux pulse avoids unwanted frequency shifts and residual coupling, which is beneficial for quantum gate operation.

As flux-tunable elements are sources of flux crosstalk in a superconducting processor, the crosstalk matrix shows the spatial effect of the flux source based on the frequency response of other probed elements~\cite{Barrett2023,Shi_2023_cal,Xiang_2023}. The diagonal elements denote the flux source, whereas the off-diagonal elements represent the flux crosstalk. Frequency-sensitive techniques like qubit spectroscopy~\cite{Barrett2023,Kosen_2024}, resonator spectroscopy~\cite{Abrams_2019,Dai_2021,Dai2024}, and Ramsey sequences~\cite{Neill2018,Sung_2021,Shi_2023_top} all characterize DC components. Phase-sensitive techniques such as Ramsey sequences~\cite{Abrams_2019,Kosen_2024} capture high-frequency components. Adding compensating pulses based on the inverse matrix suppresses these crosstalk~\cite{Barends2014,Neill2018}. Flux crosstalk compensation, along with flux pulse predistortion~\cite{Sung_2021,RLi2024,TMLi2024}, enables the prediction of idle frequencies for qubits and couplers in large quantum processors~\cite{Barends2014,Shi_2023_cal,TMLi2024}.

\begin{figure}[t]
    \begin{center}
\includegraphics[width=8.6cm]{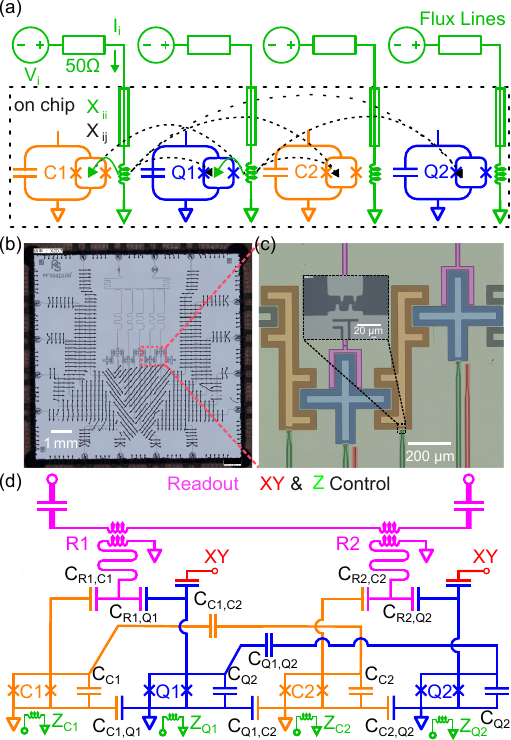}
\caption{Subsystem used to characterize magnetic flux crosstalk. (a) Illustration of flux crosstalk. Magnetic flux from the Z-lines causes unwanted frequency shifts of nearby tunable elements. (b) Optical micrograph of the chip, which has five flux-tunable transmon qubits with readout resonators and four flux-tunable couplers. (c) Optical micrograph of the subsystem consisting of two qubits (Q1, Q2) in blue and two couplers (C1, C2) in orange. The inset shows a zoomed-in picture of the device with visual termination of the flux lines adjacent to the SQUID. Color labels are assigned based on the corresponding components in the subsystem's lumped circuit model shown in (d).}
    \label{fig:device}
    \end{center}
\end{figure}

The transmon-coupler (TC) architecture~\cite{Yan2018,Sung_2021,Arute_2019,Stehlik_2021} offers an experimental testbed for implementing flux crosstalk protocols as its flux control relies on current through flux lines grounded near the qubit or coupler. We utilized a tunable two-qubit and two-coupler system to characterize flux crosstalk, as illustrated in \fref{fig:device}(a). The TC archtecture consists of transmon qubits, colored blue and labeled Q, and tunable transmon couplers, colored orange and labeled C; all equipped with flux lines, colored green and labeled Z. We study the interaction between the four flux lines $i\in\{\textrm{\textrm{$Z_{\textrm{Q1}}$},\textrm{$Z_{\textrm{Q2}}$}\,$Z_{\textrm{C1}}$},\textrm{$Z_{\textrm{C2}}$}\}$, and the SQUID loops of the target elements $j\in\{\textrm{Q1},\textrm{Q2},\textrm{C1},\textrm{C2}\}$. When the current $I_{i}$ generated by a voltage difference $V_i$ across a load impedance of $50\,\Omega$  passes through a target flux line antenna, $I_{i}$ induces two effects that generate flux crosstalk: 1) residual inductive coupling $L_{ij,i\neq j}$ between flux lines and SQUID loops of neighboring qubits and couplers and 2) unwanted fluxes threading through the SQUID structure due to return currents~\cite{Barrett2023, Kosen_2024}. The subsystem is part of a larger quantum processing unit (QPU), an optical micrograph image of which is shown in \fref{fig:device}(b). 

\sfrefs{fig:device}(c) and \textcolor{blue}{1}(d) provide a magnified image of the flux-tunable TC subsystem and its lumped circuit diagram, respectively, using matching color schemes. Each qubit has a microwave drive line and a readout resonator (R) coupled to a shared Purcell filter. The frequency arrangement is set such that the coupler frequency ($f_{\textrm{01,C}}$) is higher than the resonator frequency ($f_{\textrm{01,R}}$), which is higher than the qubit frequency ($f_{\textrm{01,Q}}$) ($f_{\textrm{01,C}}>f_{\textrm{R}}>f_{\textrm{01,Q}}$). In the present device, the chosen frequency arrangement enables the resonator to sense both the qubit and coupler frequencies through their respective dispersive shifts. The neighboring flux lines between the qubit and the coupler are separated by a distance of \SI{250}{\micro m} and are covered with wire bonds, which moderates flux crosstalk~\cite{Kosen_2024} and residual couplings with slot line modes~\cite{Wenner2011}, a common feature in processors with dense wiring, as shown in \sfref{fig:device}(c). The protocol is carried out on this subsystem which is located at the edge of the linear qubit chain. Details concerning the QPU fabrication and experimental setup information are provided in Sec. \textcolor{blue}{SII} of~\cite{SupMat}. 

To overcome the crosstalk challenge, we develop a protocol for characterizing flux crosstalk in a compact TC architecture that exploits residual inductive coupling between several flux-tunable elements to resolve the 0--1 transition frequencies of the probed element, named "Multi-Z-Line Control" (MZLC). The method only requires prior knowledge of the flux bias dependence of the qubit/coupler spectrum. It works even with coarsely optimized readout fidelity and applies to arbitrary flux biases of the probed element. We apply this method in compact TC with dense wiring of flux lines without a dedicated readout resonator and drive line for coupler. MZLC reveals average flux crosstalk of $27\,\pm\,19\,$\textperthousand$\,$ and $0.2\,\pm\,0.1\,$\textperthousand,  before and after the flux crosstalk matrix compensation. Flux crosstalk matrix cancellation after MZLC enables the coupler-assisted coupling strength to be symmetric with varying coupler detuning after compensation, easing up the tune-up of conditional-phase (CZ) gates. 

\begin{figure*}[tbp]
    \includegraphics[width=17.0cm]{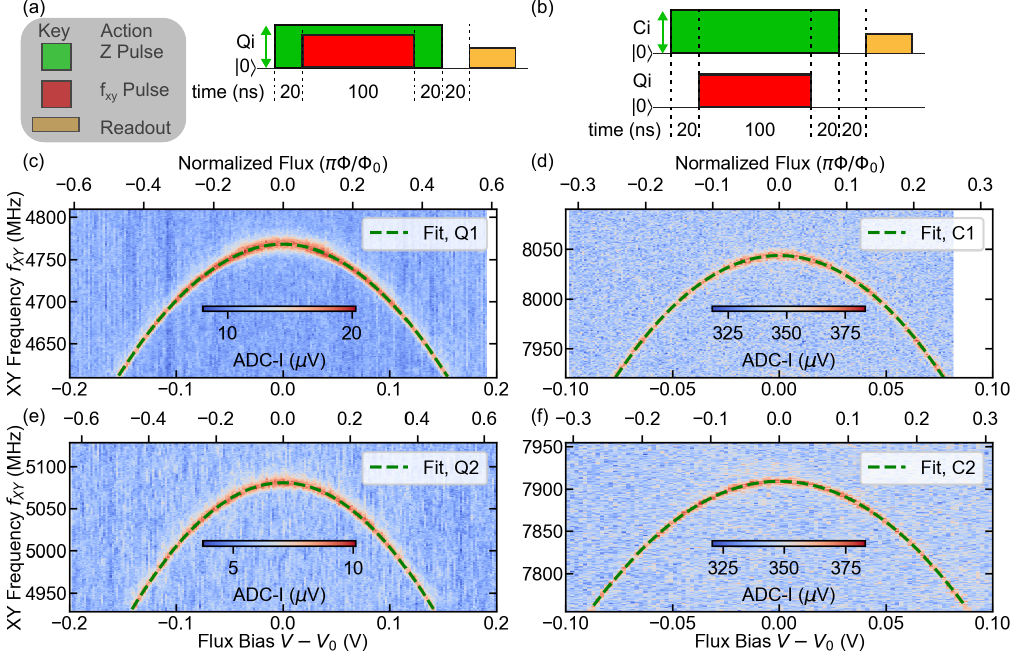}
    \caption{
    Qubit and coupler spectrum as a function of flux bias. (a) represents pulse sequences in (c) and (e) to probe qubit's f$_{\text{01}}$, using each qubit's flux bias (green), microwave drive $f_{\text{XY}}$ (red) and readout pulse (gold) as noted in the gray box. (b) represents the pulse sequences in (d) and (f) used for the coupler where the flux bias came from the coupler while the microwave and readout drives came from an adjacent qubit and resonator. Readout pulses in (a) and (c) have a \SI{20}{\nano s} delay to minimize overlap between flux pulse transients and readout pulse which degrades the contrast. We probe the qubit and coupler spectrum as a function of the flux shown in (c,e) and (d,f) for qubits and couplers, respectively. Green dashed lines refer to the fit of the f$_{\text{01}}$ transition frequencies with the flux-tunable transmon model. The top and bottom horizontal axes refer to the normalized flux bias and applied voltage, respectively.}
    \label{fig:coupler}
\end{figure*}

The MZLC protocol requires an initial knowledge of the frequency dependence of the qubits and couplers on flux bias. This information is obtained via standard two-tone spectroscopy, using the pulse schedules shown in \frefs{fig:coupler}(a) and \textcolor{blue}{2}(b), respectively, with the same color codes as \frefs{fig:device}. \sfrefs{fig:coupler}(c) and \textcolor{blue}{2}(e) reveal the frequency spectrum of the qubits Q1 and Q2, with the driving frequency $f_{\text{XY}}$ as a function of the flux bias voltage $V_{i}$. Note that every element involved in the pulse sequences shown in all figures may subject to a constant DC offset component of $V_{i}$ to compensate for the flux offset induced by trapped magnetic fields during cooldown. Any variations in the flux bias voltage amplitude shown in all pulse sequences correspond to the pulsed flux component of $V_{i}$. 

Due to the absence of dedicated readout resonators and XY lines for the couplers, we employ \textit{indirect coupler spectroscopy} (ICS) with the pulse sequence shown in \fref{fig:coupler}(b). ICS exploits the weak capacitive coupling between the coupler electrode and the XY and readout lines of the adjacent qubits to resolve $f_{01,\text{C}}$~\cite{Li_2020,collodo_2020}. In \fref{fig:coupler}(b), the XY lines of Q1 and Q2 drive the 0--1 transition of C1 and C2, respectively. Then R1 and R2 read the 0--1 transition of C1 and C2, respectively. \frefs{fig:coupler}(d) and \textcolor{blue}{2}(f) show the frequency spectra of couplers C1 and C2, respectively. In obtaining \frefs{fig:coupler}(d) and \textcolor{blue}{2}(f), the coupler is driven with elevated amplitude of microwave and readout tones to better resolve the 0--1 transition spectrum. Moreover, adding appropriate delay to overlapping drive and readout pulses enhances the contrast of the 0--1 transition in the coupler spectrum, thereby improving the signal-to-noise ratio for both qubit spectroscopy and ICS. As these weak capacitive couplings are inherent in the subsystem design, ICS is compatible with standard measurement schemes. 

Sweeping the microwave drive frequency $f_{\text{XY}}$ of the qubits' XY line with fixed XY drive and flux bias reveals a peak frequency response with a Lorentzian profile of both the qubits and couplers. The peak frequency, representing the 0--1 transition, changes with flux bias voltage $V_{i}$, and follows $f_{\text{01}}$ of the flux-tunable transmon model, which is shown as green dashed lines in \frefs{fig:coupler}(c) to \textcolor{blue}{2}(f). The model is written as ~\cite{Koch2007, Barrett2023}
\begin{equation}
\begin{split}
f_{01}(V_i)&=\left(f_{01,\max}+ E_C/h\right) \times \\  &\quad\sqrt[\leftroot{-2}\uproot{2}4]{d^2 + \left[ 1-d^{2} \right] \textrm{cos}^{2}\left(A_c \left[ V_i-V_{\text{ofs,i}}\right] \right)} \\ &\quad- E_C/h,
\end{split}
\label{eqn:f01_V}
\end{equation}
where $f_{\text{01,max}}$ is the upper sweet spot frequency, $-E_{C}/h$ is the anharmonicity, $d$ is the junction asymmetry, $A_{c}$ is a conversion factor related to the mutual inductance between the $i^{th}$ SQUID loop and the $i^{th}$ flux line, and $V_{\text{ofs,}i}$ is the offset voltage induced by the remnant flux. Notably, $A_{c}$ and $V_{\text{ofs,}i}$ are related to the normalized flux of the transmon by $A_{c}(V_{i}-V_{\text{ofs,}i})=\pi\Phi/\Phi_0$ where $\Phi$ is the total flux that traverses the SQUID loop and $\Phi_0$ is the flux quanta. The transmon anharmonicity is determined by two-photon excitation measurements. The junction asymmetries for both qubits and couplers are based on the nominal design in Sec. \textcolor{blue}{SI} of~\cite{SupMat}. Hence, $A_{c}$, $f_{\text{01,max}}$ and $V_{\text{ofs,}i}$ are extracted when fitting the model with the driven spectra in \frefs{fig:coupler}(c) and \textcolor{blue}{2}(f), expressed in voltages from the analog-to-digital converter in-phase channel (ADC-I)~\cite{Sank2025}. These parameters map the idle transmon frequency to its corresponding flux bias ~\cite{Koch2007}, and vice versa after flux crosstalk compensation (see Sec. \textcolor{blue}{SIII A} in~\cite{SupMat} for details).

\begin{figure}[!htbp]
    \includegraphics[width=8.5cm]{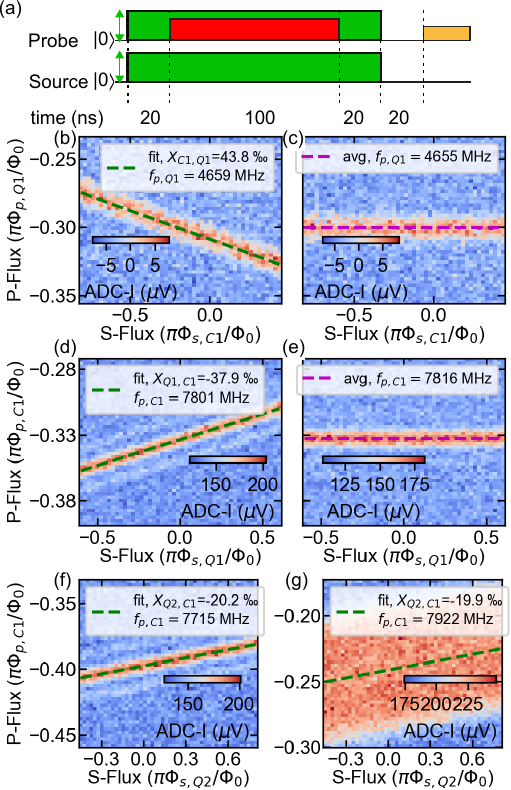}
    \caption{Characterization of flux crosstalk. (a) Pulse sequence used to extract crosstalk of MZLC: a probing element is driven at a fixed frequency, while both probe voltage and source voltage are applied simultaneously. The XY, Z, and readout pulses have \SI{20}{\nano s} delays to avoid timing mismatch artifacts. Spectra with crosstalk are illustrated in intensity plots with C1 as source and Q1 as probe (b) and vice versa (d). The horizontal, vertical, and intensity axes are the source flux (S-Flux), probe flux (P-Flux), and readout signal voltage, respectively. Applying flux-compensated pulses to (b) ($-0.39\times\text{P-Flux}_\text{Q1}$) and (d) ($-0.34\times\text{P-Flux}_\text{C1}$) after characterizing the flux crosstalk results in spectra immune to S-Flux as shown in (c) ($-0.39\times\text{P-Flux}_\text{C1}$) and (e) ($-0.33\times\text{P-Flux}_\text{Q1}$), respectively. In (f) ($-0.40\times\text{P-Flux}_\text{C1}$) and (g) ( $-0.24\times\text{P-Flux}_\text{C1}$), as the frequency detuning of probe C1 from its sweet spot frequency of $f_\text{C1,max}(\Phi_{p}=0)=8046.6\,\pm\,0.2\,$MHz decreases, the spectral linewidth of C1 increases. All measurement conditions in this figure have an identical readout pulse buffer and initialization time.} 
    \label{fig:crosstalk}
\end{figure}

We then investigate the flux crosstalk for the qubits and couplers. \sfrefs{fig:crosstalk}(a) shows the pulse sequence of the MZLC protocol with a similar legend in \fref{fig:coupler}(a). Before the protocol begins, the XY pulses for both the probe and source elements are calibrated at the maximum transmon frequency. Next, the DC flux biases of the probe and source are set to their respective target operating points. The XY pulse frequency is then fixed to match the 0–-1 transition at the target operating point. To characterize the flux crosstalk, we simultaneously sweep the probe flux pulse amplitude (P-Flux) and the source flux pulse amplitude (S-Flux), ensuring that both pulses have the same duration. To prevent timing mismatch artifacts, a \SI{20}{\nano s} buffer time is inserted between the flux, XY, and readout pulses. After the XY pulse is turned off, an additional 20 ns buffer is applied before both the S-Flux and P-Flux pulses are diabatically turned off. A \SI{20}{\nano s} delay is also introduced between the flux and readout pulses to reduce transient effects from the flux pulses. Finally, a readout pulse from the probe is used to resolve the probe frequency.

\begin{figure*}[t]
    \includegraphics[width=17.0cm]{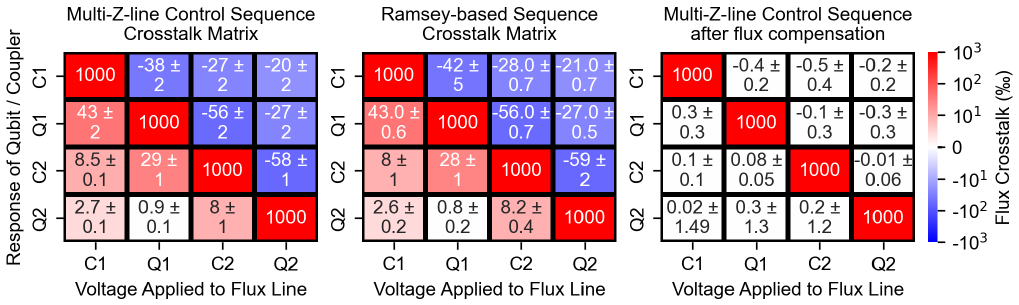}
    \caption{
    Flux crosstalk matrices between tunable elements. Rows correspond to detector elements, and columns to source Z-lines. (a) Displaying results obtained from two-tone spectroscopy averaged over 100 repetitions to estimate statistical uncertainty. (b) shows corresponding results from Ramsey, which agrees with statistical error to (a). (c) show the crosstalk matrix after compensation, demonstrating suppression of all off-diagonal elements to below $0.5\,$\textperthousand. Values smaller than their uncertainties are kept to show the nominal measurement mean, which otherwise rounds to zero.}
    \label{fig:table_mat}
\end{figure*}

In principle, an arbitrary flux pulse duration can work for MZLC. To maintain consistency and avoid measurement-induced dephasing of the qubit due to residual photons inside the cavity after readout, we set the initialization time to \SI{200}{\micro s}, which is more than twice the minimum time of the resonator R1 with a long passive reset time of $2\pi\times10\kappa^{-1}\,$ of about \SI{92}{\micro s}, where $\kappa$ is the resonator decay rate (see Table \textcolor{blue}{S2} from~\cite{SupMat} for details). We also set both the flux pulse duration and the microwave drive pulse duration to \SI{100}{\nano s} throughout the study (see Sec. \textcolor{blue}{SIII F} of \cite{SupMat} for pulse duration dependence). Therefore, we can ensure that the crosstalk $X_{ik}$, where $i  \text{ and } k$ are flux lines and $i \neq k$, is not affected by variations in the pulse configuration.

\sfrefs{fig:crosstalk}(b) and \textcolor{blue}{3}(d) show crosstalk measurement results using MZLC. The horizontal and vertical axes represent the applied voltage on the source and probe flux lines, respectively, which are converted to flux quanta, named S-Flux ($\pi \Phi_{s,i}/\Phi_0$) and P-flux ($\pi \Phi_{p,i}/\Phi_0$). Each data point represents a frequency response ($f_\text{01}$) of the qubits (couplers) as a detector under a fixed XY frequency, driven at the nominal qubit frequency extrapolated from the fit in \frefs{fig:coupler}(c) and \textcolor{blue}{2}(d), and a simultaneous bias of the source and probe elements. 

MZLC resolves the behavior of the 0--1 transition of the probe by monitoring changes in its population, which is between $|0\rangle$ and $|1\rangle$ for a weak pulsed XY drive, with different combinations of P-Flux and S-Flux. The combined pulses will allow additional sampling of the flux crosstalk effect with better precision, and can ensure the linear correlation between probe and source elements,  instead of 1D scans as done in other works~\cite{Neill2018}. The scheme hinges on detecting flux crosstalk by observing undesirable frequency shifts on the detector induced by S-Flux, which is visually distinct from other sources of unwanted frequency shifts. In the absence of flux crosstalk, the 0--1 transition of the probe has a characteristic P-Flux that is unaffected by the S-Flux sweep. With flux crosstalk, the 0--1 transition appears to vary linearly with the S-Flux sweep. The magnitude of the slope captures the total applied flux magnitudes as done previously with other techniques~\cite{Neill2018}. Other sources of frequency shift, such as dispersive shifts caused by neighboring flux-tunable elements, and ZZ interaction-induced shift, appear as a nonlinear dependence of the probe's 0--1 transition~\cite{Barrett2023} at different combinations of simultaneously biased P-Flux and S-Flux.

It is worth noting that an alternative method is the Ramsey-based method~\cite{Neill2018,Sung_2021,Shi_2023_top}, which observes the same sign and magnitude of crosstalk, as shown in \fref{fig:table_mat}(b) and Fig. \textcolor{blue}{S2} of the supplemental material~\cite{SupMat}. 
Within the S-flux ranges shown in \frefs{fig:crosstalk}(b) and \textcolor{blue}{3}(d), the $f_{\text{01}}$ of the probe elements are linearly proportional to the flux quanta of the source element, implying linear crosstalk between neighboring flux-controlled elements. We model this linear dependence on the source frequency $f_{s,i}$ as
\begin{equation}
f_{p,k}=X_{ik}f_{s,i}+f_{p,i}=\frac{-\Delta V_{p,k}}{\Delta V_{s,i}}f_{s,i} + f_{p,i},
\label{eqn:fp_fs}
\end{equation}
where $X_{ik}=-\Delta V_{p,k}/\Delta V_{s,i}$ is the flux crosstalk coefficient obtained from the slope of \frefs{fig:crosstalk}(b) and \textcolor{blue}{3}(d). We label $X_{ik}$ as the non-diagonal elements of the flux crosstalk matrix $X$ along with its normalized diagonal term ($X_{ii}=1$). With $X$ measured, one can calculate the inverse matrix, which leads to a cancellation matrix $X^{-1}$. After applying $X^{-1}$ to reconfigure our flux pulses, the remeasured flux crosstalk maps in \frefs{fig:crosstalk}(c) and \textcolor{blue}{3}(e) show that the S-Flux does not affect on the transition frequency of the probe element. The results not only support the validity of \eref{eqn:fp_fs} but also confirm a significant suppression of flux crosstalk. Notably, the intercept $f_{p,i}$ extracted from \frefs{fig:crosstalk}(b) and \textcolor{blue}{3}(d) is a good estimate of the probe idle frequency $f_{p,k}$ after flux compensation, as shown in \frefs{fig:crosstalk}(c) and \textcolor{blue}{3}(e). We attribute these megahertz discrepancies to uncorrected long-time-scale transients of the flux pulses~\cite{Sung_2021,TMLi2024,RLi2024} (see Fig. \textcolor{blue}{S4} and Sec. \textcolor{blue}{SIII F} of~\cite{SupMat} for details). 

These megahertz discrepancies, if not corrected first with flux predistortion and then flux crosstalk compensation, lead to coherent errors that affect the 1Q gate fidelities of qubits in a multi-qubit processor having simultaneous flux pulse biasing (see Secs. \textcolor{blue}{SVI and SVII A} of~\cite{SupMat} for details).

When using the MZLC approach, detuning the probe element away from its upper sweet spot leads to a narrow linewidth, which improves the precision of flux crosstalk measurement, as shown in \fref{fig:crosstalk}(f). Conversely, detuning the probe element towards the upper sweet spot frequencies, as shown in \fref{fig:crosstalk}(g), results in a broadened spectral linewidth and larger error bar in the crosstalk fitting. These linewidths are consistent with the qubit and coupler spectra in \fref{fig:coupler}, which are observed in previous works~\cite{Schuster_2005,Paik2011,Whittaker2014,Sung_2021,Anferov_2024}.

In fast-flux qubit spectroscopy, whose pulse time is much faster than the relaxation and dephasing time $t\ll T_1,T_2^*$, the linewidth of the qubit spectrum is Fourier-limited by the inverse of the pulse duration time. However, the XY Rabi drive, which is proportional to the transmon's dipole matrix element, and consequently the fourth root of the Josephson energy~\cite{Sung_2022}, has periodic dependence on the applied flux. The flux-dependent XY Rabi drive modulates the effective qubit linewidth. As our detection undergoes repeated ensemble averaging per pixel to resolve the 0--1 transition, low-frequency drifts (like qubit frequency drifts) are averaged out, leading to qubit dephasing~\cite{Martinis_2003}. At a weak XY drive, the observed pulsed spectroscopy lineshape becomes Gaussian-like due to inhomogeneous broadening~\cite{Blais_2021,Martinis_2003,Krantz_2019}. Despite the observed broadening, peak thresholding and curve fitting improve the frequency shift resolution (see Sec.~\textcolor{blue}{SV} of~\cite{SupMat} for details). Nonetheless, the sign and amplitude of the crosstalk in \frefs{fig:crosstalk}(f) and \textcolor{blue}{3}(g) remain unchanged.      

\begin{table}[htbp] 
\caption{
Metrics for Flux Crosstalk $X_{ik}$ (in \textperthousand) before and after flux crosstalk compensation based on \frefs{fig:table_mat}(a) and \textcolor{blue}{4}(c).} 
\begin{ruledtabular}
\begin{tabular}{cccc}
\textrm{Metric} &
\textrm{Notation} &
\textrm{Before} &
\textrm{After} \\
\colrule
Largest Negative & $\text{Min}(X_{ik})$& $-58\,\pm\,1$ & $-0.5\,\pm\,0.4$ \\
Largest Positive & $\text{Max}(X_{ik})$ & $43\,\pm\,2$ & $0\,\pm\,1$ \\
Average & $\langle |X_{ik}| \rangle_{i \ne k}$ & $27\,\pm\,19$ & $0.2\,\pm\,0.1$ \\
Total & $\sum_{i \ne k}|X_{ik}|$ & 318.4 & 2.42 \\
\makecell{Matrix \\ Asymmetry} & \makecell{$\sum_{i \ne k}$\\$||X_{ik}|-|X_{ki}||$} & 143.6 & 0.94
\end{tabular}
\end{ruledtabular}
\label{tab:X_stats}
\end{table}

We summarize the crosstalk matrices measured using the MZLC and Ramsey sequences in \frefs{fig:table_mat}(a) and \textcolor{blue}{4}(b), respectively. Both methods yield consistent results within the statistical error. We discuss the efficiency of MZLC and Ramsey schemes in Secs. \textcolor{blue}{SIII B, and SVII} of~\cite{SupMat}. Table \ref{tab:X_stats} shows the statistical properties of $X_{ik}$ before and after flux compensation in \frefs{fig:table_mat}(a) and 4(c), respectively. Few observations of the crosstalk matrix are as follows. First,  the average uncompensated crosstalk has the same order as its extreme values (Fig. \textcolor{blue}{S3} of~\cite{SupMat} presents the same results in dB). Its total crosstalk is similar to previous flux-tunable two-qubit transmons~\cite{Chow2010}. Its matrix asymmetry exhibits an imbalance likely stemming from the leftward routing layout of our flux lines, producing asymmetric inductive coupling and return current. A careful routing of flux lines~\cite{Kosen_2024} with the return current flowing directly on the PCB ground~\cite{Niu2024} could passively reduce the flux crosstalk and matrix asymmetry to a certain extent. Lastly, these metrics dropped one to two orders of magnitude after flux compensation; residual crosstalk averages near zero, with a standard deviation of a similar scale. The flux-compensated crosstalk becomes decoupled, uniform, and reciprocal. 

\begin{figure}[!tbp]
    \includegraphics[width=8.5cm]{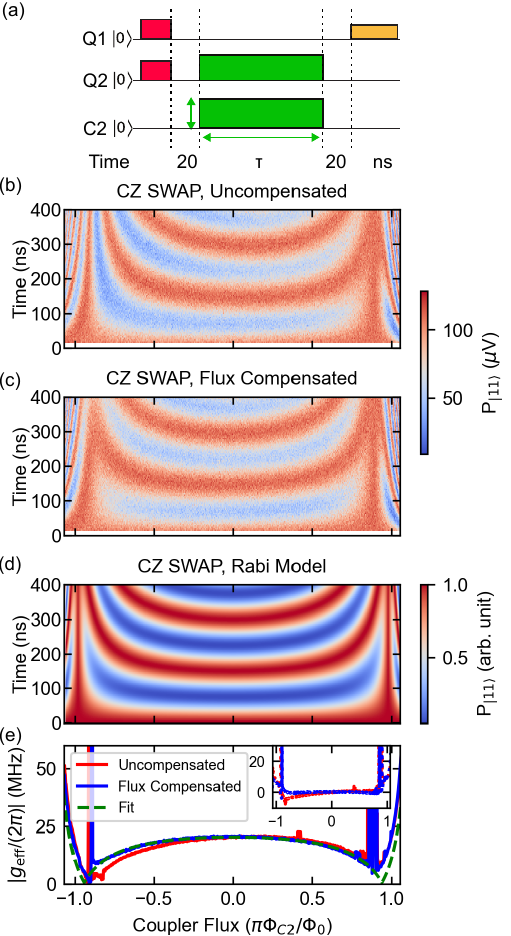}
    \caption{
    Time-domain measurements of the coherent oscillations between $|11\rangle$ (red) and $|02\rangle$ (blue) states, named CZ SWAP, under varying coupler flux and different flux pulse durations. The pulse schedule in (a), with the XY drive being the $\pi_{01}$ pulse, and the results before flux compensation (b) and after flux compensation (c) are displayed. The flux-compensated time-domain measurements agree with simulations in (d). (e) 2D Fast Fourier transform (FFT) of the time-domain oscillations in (b), (c), and (d), which quantifies the conditional phase-mediated coupling strength $|g_{\mathrm{eff}}|$ as a function of coupler flux. Inset shows the residuals (dotted lines) expressed in MHz. Panels from (b) to (e) share the same x-axis, representing the applied flux on the coupler.}
    \label{fig:CZ_bridge}
\end{figure}

In addition to compensating for flux crosstalk, we applied this method to help locate the optimal bias point for conditional phase (CZ) gate operations~\cite{Foxen_2020,Sung_2021,RLi2024}. In TC architectures, the coupler-mediated CZ gate works by tuning the frequencies of the qubits so that the states $\vert11\rangle$ and $\vert20\rangle$ become resonant, resulting in a state swap. The effective coupling strength $g_{\mathrm{eff}}$ in this regime is susceptible to detuning between the frequencies of qubits and couplers, which in turn depends on the applied flux bias.

We show measured coupler-tuned CZ SWAP experiments using the pulse schedule as shown in \fref{fig:CZ_bridge}(a) (see Sec. \textcolor{blue}{SIII D} of~\cite{SupMat} for the relevant CZ SWAP idle frequencies). The behavior seen in \frefs{fig:CZ_bridge}(b) and \textcolor{blue}{5}(c) before and after flux compensation, respectively, are compared with \fref{fig:CZ_bridge}(d), which is derived from the Rabi oscillation for 2Q gates~\cite{Anferov_2024,hellings_2025}, accounting for the finite rising time of the diabatic flux pulse. This finite rise time—an effective flux transient representing the combined, rather than individual, components of flux pulse distortions—results in fewer Rabi oscillation cycles and, consequently, a reduced interaction time for two-qubit (2Q) gates (see additional details of the model from Secs. \textcolor{blue}{SIII D to SIII F} of~\cite{SupMat}). Due to phase errors arising from the uncalibrated pulse distortions in flux biasing~\cite{Shi_2023_cal,RLi2024}, \frefs{fig:CZ_bridge}(b) and \textcolor{blue}{5}(c) may show a small deviation from \fref{fig:CZ_bridge}(d). 

As finite resolution in \frefs{fig:CZ_bridge}(b) and \textcolor{blue}{5}(c) prevent us to extract the coupling strength via FFT like in other works~\cite{Sete_2021,Liang_2023_swap}, we fit the readout signals representing the $\ket{11}$ population versus time duration in \frefs{fig:CZ_bridge}(b) and \textcolor{blue}{5}(c) with a cosine function to extract the effective coupling strengths $\Omega=\sqrt{4g_{\mathrm{eff}}^{2}+\delta_{21}^{2}}$, where $g_{\mathrm{eff}}$ is the coupler-mediated coupling strength for the CZ gate, $\delta_{21}$ is the detuning between the $|11\rangle$ and $|02\rangle$ states. Using the extracted qubit and coupler parameters in another chip with identical circuit design using \eref{eqn:f01_V}, we determined $\delta_{21}/2\pi=9.33\,$MHz. After subtracting the CZ detuning from $\Omega$, we extracted $|g_{\text{eff}}|$ as a function of the flux of the coupler $C_{2}$, shown in \fref{fig:CZ_bridge}(e), for both compensated and uncompensated cases.    

The CZ coupling strength in a TC architecture (Q1-C2-Q2) is written as~\cite{Sete_2021,Liang_2023_swap} 
\begin{equation}
g_{\text{eff}} = \sqrt{2}\left(g_{12} - \frac{g_{1c}g_{2c}}{2}B_{02} \right),
\label{eqn:g_cz1}
\end{equation}
where $g_{12}$ is the direct coupling between Q1 and Q2, $g_{1c}$ is the coupling strength between Q1 and coupler C2, $g_{2c}$ is the coupling strength between Q2 and coupler C2, and
\begin{equation}
B_{02} = \frac{1}{\Delta_1} + \frac{1}{\Delta_2-E_{C,\text{Q2}}/\hbar} + \frac{1}{\sum_1} + \frac{1}{\sum_2 + E_{C,\text{Q2}}/\hbar},
\label{eqn:g_cz2}
\end{equation}
where $\Delta_{1}/2\pi=f_{01,\text{C2}}-f_{01,\text{Q1}}$ and $\Delta_{2}/2\pi=f_{01,\text{C2}}-f_{01,\text{Q2}}$ are the frequency detunings between the coupler and the qubits, and $\sum_{1}/2\pi=f_{01,\text{C2}}+f_{01,\text{Q1}}$ and $\sum_{2}/2\pi=f_{01,\text{C2}}+f_{01,\text{Q2}}$ are the frequency sums of the transition frequencies. Moreover, $-E_{C,\text{Q1}}/\hbar$ and $-E_{C,\text{Q2}}/\hbar$ are the anharmonicities of Q1 and Q2, respectively. To fit the flux-compensated $|g_{\text{eff}}|$ in \fref{fig:CZ_bridge}(e), we use Eqs.~\eqref{eqn:g_cz1}--\eqref{eqn:g_cz2} with an added coupling strength of $|435|\,$kHz, which could be reduced through better readout fidelity and rigorous flux predistortions~\cite{RLi2024,Shi_2023_cal,hellings_2025}. The inset in \fref{fig:CZ_bridge}(e) accentuates the effect of flux compensation on predicting $|g_{\text{eff}}|$. At $|\Phi_\text{C2}|<0.88\Phi_0$, uncompensated flux pulses generate asymmetric $|g_{\text{eff}}|$ with megahertz offsets from the fit, whereas compensated flux pulses produce predictable $|g_{\text{eff}}|$.

Non-idealities in \fref{fig:CZ_bridge}(e) and its inset become evident when comparing the measured $|g_{\text{eff}}|$ with the fit at $|\Phi_{\text{C2}}|\geq 0.88\Phi_0$. The fit cannot explain the sharp peaks at $|g_{\text{eff}}/2\pi|$ at $|\Phi_\text{C2}|\approx0.88\Phi_0$ ($f_{\text{C2}}\approx$ \SI{5.94}{\giga Hz}) as the predicted singularities occur beyond the range of the coupler flux displayed in \fref{fig:CZ_bridge}(e)~\cite{Liang_2023_swap}. The singularities probably come from the strong coupling between the readout resonator R2 and the coupler C2 as $f_{\text{C2}}\approx f_{\text{R2}}$ (see Table \textcolor{blue}{S2} of~\cite{SupMat}), which produces hybridized modes as exploited in other works~\cite{chen2024_reset,zhang2025_aswap}. Hybridization of modes adds difficulty in nulling $|g_{\text{eff}}|$, which explains the monotonic increase in the two residuals when $|\Phi_\text{C2}|>0.88\Phi_{0}$. 

The findings in \frefs{fig:CZ_bridge} imply that flux crosstalk compensation creates a magnetic flux crosstalk-free \uline{intuitive} digital twin of the coupler-mediated CZ gate. This twin is based on empirical adjustments of the Rabi oscillation model~\cite{hellings_2025}, whose analytical form is validated by carefully calibrated two-qubit state measurements~\cite{Krizan_2025} and initially modelled as three-interacting qutrits~\cite{Sung_2021}. This digital twin helps identify non-idealities in two-qubit gates such as reduced 2Q Rabi oscillation cycles brought upon by an effective flux transient~\cite{Sung_2021,TMLi2024,hellings_2025} and unintended hybridization of the qubit and coupler modes. Shaping the flux pulse to account for flux transients~\cite{Sung_2021,TMLi2024,hellings_2025} while avoiding frequencies that invoke unwanted hybridization of modes~\cite{chen2024_reset} could help to match the digital twin with experimental results. We consider the proposed measurement, modeling, and analysis protocol to be suitable for scaling to hundreds of qubits with the aid of high-performance computing~\cite{Mohseni_2025} (see Sec. \textcolor{blue}{SVII} of~\cite{SupMat} for supporting details).

In this work, we develop the MZLC protocol for characterizing flux crosstalk in a TC architecture. The method exploits residual inductive coupling between several flux-tunable elements to resolve the 0--1 transition frequencies of the probed element. Whereas other flux crosstalk characterization strategies use Rabi spectroscopy ~\cite{Barrett2023,Kosen_2024} and Ramsey sequences on a single-biased element to characterize the flux crosstalk matrix, both rely on good readout fidelity of the qubit/coupler. MZLC has the advantage of characterizing flux crosstalk even in modest readout SNR and an arbitrary idle flux bias range. The method requires knowledge of the dependence of the qubit (coupler) spectrum on flux bias, which is routine in the bring-up of superconducting processors. As discussed in the supplemental material~\cite{SupMat}, its primary advantage stems from avoiding the prerequisite calibration steps for precise $\pi$ and $\pi/2$ gate operations when calibrating flux crosstalk. This combination makes the MZLC a simple, scalable, robust and low-overhead tool for characterizing the flux crosstalk matrix of a transmon-coupler subsystem and optimizing CZ gate operation points. Lastly, compared with the CZ SWAP model, the experimental result shows a nearly identical map with zero crosstalk. Therefore, it provides a way to create a digital twin for the TC system CZ optimization.

\section*{Supplemental Material}
The supplemental material~\cite{SupMat} includes the following: I) details of the QPU fabrication and design; II) experimental details and essential QPU parameters; III) calibration between flux and qubit-coupler interactions; IV) flux-induced inhomogeneous effects on linewidth in MZLC; V) spectroscopic signal-to-noise ratio and frequency-shift uncertainty of MZLC; VI) effect of flux crosstalk compensation on single-qubit gate fidelities; VII) scalability of flux crosstalk characterization and compensation for large-scale superconducting processors. 

\section*{Data Availability}
The data that support the most of the findings of this study are openly available in Zenodo at http://doi.org/10.5281/zenodo.17639715. Check the latest version for the updated datasets. All other data are available from the corresponding authors upon request.

\section*{Acknowledgment}
We acknowledge funding support from the Academia Sinica Grand Challenge project (AS-GCP-112-M01), Grand Challenge Program Seed Grant (AS-GCS-114-M04) National Quantum Initiative (AS-KPQ-111-TQRB), and NSTC (113-2119-M-001-008). C.-T. K. acknowledges the funding support from NSTC 2030 Cross-Generation Young Scholars(112-2628-M-001-004). H.-S. Goan acknowledges support from the National Science and Technology Council (NSTC), Taiwan, under Grants No. NSTC 113-2112-M-002-022-MY3, and No. 114-2119-M-002-018, and from the National Taiwan University under Grants No. NTU-CC-114L8950, and No. NTU-CC114L895004.  

\section*{Author Declaration}
\subsection*{Conflict of Interest}
The authors have no conflicts to declare.

\subsection*{Author Contributions}
C.-D. C. and C.-T. K. conceived the devices and supervised the project with contributions from H.-S. G., S.-Y. H. and C.-H. W.. C.-C. C. and J.-Y. W. designed the device layout with additional inputs from T.-H. L. Y.-C. C. coordinated the device fabrication. Y.-C. C., C.-S. W., C.-D. C., and C.-T. K. provided input on the fabrication methodology. N.-Y. L., L.-C. H. and Y.-S. H. conducted the experiment. M.A.C.A., N.-Y. L., C.-H. M., L.-C. H., and C.-T. K. analyzed the data and wrote the manuscript; all authors discussed the results and contributed to the manuscript. 

\bibliographystyle{apsrev4-2}
\bibliography{main} 

\end{document}


\preprint{APS/123-QED}

\title{Supplemental Material for "Characterizing and Mitigating Flux Crosstalk in Superconducting Qubits-Couplers System"}

\author{Myrron Albert Callera Aguila}
\thanks{M. A. C. A., N.-Y. L. and C.-H. M. contributed equally in the manuscript}
\affiliation{Research Center for Critical Issues, Academia Sinica, Guiren, Tainan, 711010, Taiwan}

\author{Nien-Yu Li}
\thanks{M. A. C. A., N.-Y. L. and C.-H. M. contributed equally in the manuscript}
\affiliation{Institute of Physics, Academia Sinica, Nankang, Taipei, 11529, Taiwan}
\affiliation{Department of Physics, National Taiwan University, Da'an District, Taipei 10617, Taiwan}

\author{Chen-Hsun Ma}
\thanks{M. A. C. A., N.-Y. L. and C.-H. M. contributed equally in the manuscript}
\affiliation{Institute of Physics, Academia Sinica, Nankang, Taipei, 11529, Taiwan}
\affiliation{Department of Physics, National Taiwan University, Da'an District, Taipei 10617, Taiwan}

\author{Li-Chieh Hsiao}
\affiliation{Research Center for Critical Issues, Academia Sinica, Guiren, Tainan, 711010, Taiwan}

\author{Yi-Shiang Huang}
\affiliation{Institute of Physics, Academia Sinica, Nankang, Taipei, 11529, Taiwan}

\author{Yen-Chun Chen}
\affiliation{Research Center for Critical Issues, Academia Sinica, Guiren, Tainan, 711010, Taiwan}

\author{Teik-Hui Lee}
\affiliation{Research Center for Critical Issues, Academia Sinica, Guiren, Tainan, 711010, Taiwan}

\author{Chin-Chia Chang}
\affiliation{Institute of Physics, Academia Sinica, Nankang, Taipei, 11529, Taiwan}

\author{Jyh-Yang Wang}
\affiliation{Department of Electrical Engineering, Feng Chia University, Xitun, Taichung 407301, Taiwan}

\author{Ssu-Yen Huang}
\affiliation{Department of Physics, National Taiwan University, Da'an District, Taipei 10617, Taiwan}

\author{Hsi-Sheng Goan}
\affiliation{Department of Physics, National Taiwan University, Da'an District, Taipei 10617, Taiwan}
\affiliation{Center for Quantum Science and Engineering, National Taiwan University, Taipei 106319, Taiwan}
\affiliation{Physics Division, National Center for Theoretical Sciences, Taipei 106319, Taiwan}

\author{Chiao-Hsuan Wang}
\affiliation{Department of Physics, National Taiwan University, Da'an District, Taipei 10617, Taiwan}
\affiliation{Center for Quantum Science and Engineering, National Taiwan University, Taipei 106319, Taiwan}
\affiliation{Physics Division, National Center for Theoretical Sciences, Taipei 106319, Taiwan}

\author{Cen-Shawn Wu}
\affiliation{Department of Physics, National Changhua University of Education, Changhua, Changhua 500207, Taiwan}
\affiliation{Research Center for Critical Issues, Academia Sinica, Guiren, Tainan, 711010, Taiwan}

\author{Chii-Dong Chen}
\affiliation{Institute of Physics, Academia Sinica, Nankang, Taipei, 11529, Taiwan}
\affiliation{Research Center for Critical Issues, Academia Sinica, Guiren, Tainan, 711010, Taiwan}

\author{Chung-Ting Ke}
\affiliation{Institute of Physics, Academia Sinica, Nankang, Taipei, 11529, Taiwan}
\affiliation{Research Center for Critical Issues, Academia Sinica, Guiren, Tainan, 711010, Taiwan}

\date{\today}
             
\maketitle

\section{\label{sec:sup_fab}Device Fundamentals} 

We fabricated the device on a high-resistivity (100) Si substrate ($\rho=10\,\textrm{k$\Omega$\,cm}$) based on a 4-inch wafer process. We first performed a UV ozone clean-
ing process for two minutes, followed by a hydrofluoric acid "HF" dip to remove native oxides on the substrate for one minute, and then rinsing with deionized water (DIW). The N$_{2}$-dried wafer is baked at 200 \textdegree C
in a vacuum before spin coating with electron beam resist. A 100 kV e-beam writer defines all circuit elements prior to resist development. We then transfer the patterned wafer to conduct conventional double-angle evaporation, forming a Manhattan junction as well as other superconducting regions. The thickness of Al is 50 nm and 100 nm for the first and second layers, respectively. Lastly, we do a gentle resist lift-off and cleaning process to remove the resist, metal and organic residues. An automatic wire-bonder is used to tie the ground planes on the chip to reduce flux crosstalk~\cite{Wenner2011}.

\begin{table}[htbp]
\caption{
Simulated Parameters for Qubits and Couplers in the CZ idle configuration ($f_{R}=6.0\,$GHz, $f_{\text{Q1,idle}}=4720\,$MHz, $f_{\text{Q2,idle}}=4931\,$MHz, $f_{\text{C1,idle}}=7735\,$MHz and $f_{\text{C2,idle}}=5379\,$MHz) based on Ansys Q3D simulations}
\begin{ruledtabular}
\begin{tabular}{ccccc}
\textrm{Transmon}&
\textrm{C1}&
\textrm{C2}&
\textrm{Q1}&
\textrm{Q2}\\
\colrule
$E_{C}/h\,$(MHz) & 153 & 153 & 206 & 206 \\
$g_{qr}/2\pi$ (MHz) & 8.10 & 13.4 & 106 & 111 \\
$g_{\text{R1,R2}}/2\pi$ (MHz) & - & - & 1.98 & 1.98 \\
\end{tabular}
\end{ruledtabular}
\label{tab:q_design}
\end{table}

The e-beam pattern layout is drawn using the KLayout software. Ansys Q3D is used to calculate the self-capacitance and coupling capacitance of the circuit layout. These capacitances translate into target charging energies and resonator-qubit coupling strengths based on previous works\cite{Sung_2022,Anferov_2024}, as listed in Table~\ref{tab:q_design}. Notably, the calculation of the resonator-resonator coupling strength serves as the weakest coupling magnitude in the subsystem.     

The lithographic areas of the SQUIDs for qubits and couplers have distinct designs. The square junction areas of the SQUIDs of our qubits are nominally asymmetric ($A_{Q,j1}=0.07\,\times\,0.07\,\mu$m$^{2}$, $A_{Q,j2}=0.152\,\times\,0.152\,\mu$m$^{2}$). The square junction areas for the couplers are nominally symmetric, although with different widths ($A_{C,j1}=A_{C,j2}=0.228\,\times\,0.228\,\mu$m). 
The Josephson energy is written as
\begin{equation}
    E_J= \frac{\Phi_0\Delta}{4eR_{N}}\textrm{tanh}\frac{\Delta}{2k_{\textrm{B}}T},
    \label{eqn:sEj_RN}
\end{equation}
where $\Phi_0$ is the magnetic flux quantum, $e$ is the electron charge, $k_{B}T$ is the thermal energy, and $\Delta$ is the superconducting energy gap of aluminum. We define the normal state resistance $R_{N}=R_{J}/A$~\cite{Osman_2021}, where $R_{J}$ is the junction resistance per unit area and $A$ is the junction area. Hence, we get a relation of $E_{J}\propto A$. For asymmetric junctions, the junction asymmetry $d$ of the transmons is written as
\begin{equation}
    d= \frac{\gamma-1}{\gamma+1},
    \label{eqn:sd_coup}
\end{equation}
where $\gamma=E_{j2}/E_{j1}$. Based on the designed areas of the junctions in the qubits, $\gamma_{\text{Q}}=A_{\text{Q},j2}/A_{\text{Q},j1}=4.715$, and the junction asymmetry of the qubits is $d_\text{Q}=0.65$. For symmetric junctions of the coupler SQUIDs, $d_\text{C}=0$. These values of $d$ are integral in confirming the parameters of the qubit and coupler spectra, as noted in the latter section and in the main text.

\section{\label{sec:sup_meas_setup}Measurement Wiring and Experimental Tuneup}

\begin{figure}[htbp]
    \centering
    \includegraphics[width=8.6cm]{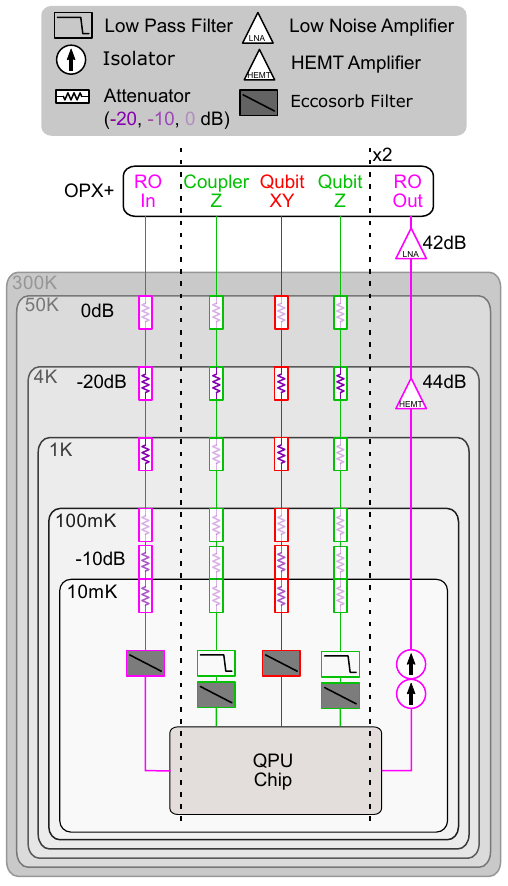}
    \caption{Wiring schematic of the measurement setup. Different resistor colors in the attenuator refer to the attenuation level (see legend). Colored lines refer to the qubit readout (RO, magenta), microwave drive (XY, red) and flux (Z, green) control noted in Figs. \textcolor{blue}{1}(c,d) of the main text. Dashed lines refers to region in the measurement setup that has duplicated wiring and control for multiple qubits and couplers.}
    \label{sfig:wiring}
\end{figure}

The chip, which contains the quantum processing unit (QPU), has five qubits and four Transmon Couplers (TC) that are arranged in a linear array. Each qubit has its dedicated $\lambda/4$ coplanar waveguide (CPW) resonators, with a shared $\lambda/2$ CPW to ensure fast readout (RO). Typical readout resonator frequencies are around $5.9-6.1\,$GHz. The chip is mounted on a single-layer tin-plated PCB board with a small sliver of GE varnish on the bottom part of the chip to provide thermal link with the base temperature of the dilution fridge (DR). Aluminum wire bonds connect the signal and readout lines of the chip to the $50\,$ohm electrodes of the PCB board~\cite{LiuCH_2023}. The PCB board is covered with an aluminum cap followed by a mu-metal cap to attenuate the magnitude of trapped flux~\cite{Arute_2019}. 

The chip mount is tightly screwed onto an Au-plated copper plate for thermalization of the Bluefors LD-400 DR, as shown in \fref{sfig:wiring}.  The 0-dB attenuators thermally connect the inner conductor of the cables to different stage plates of the DR. Both readout (RO-in) and microwave drive (Qubit-XY) lines have similar attenuators to minimize the population of the qubit with thermal photons~\cite{Krinner_2019}. Superconducting coaxial cables connect the SMA connectors between \SI{4}{K} and \SI{10}{\milli K} for the flux lines (Coupler Z and Qubit Z), and output readout lines (RO-out), as shown in \fref{sfig:wiring}. A dissipative 500-MHz low-pass filter (LPF, Mini Circuit VLFX-500+) is used to damp high-frequency standing waves that may be created when applying baseband pulses between the flux line antenna and the low-pass filter~\cite{Arute_2019}. Low loss Eccosorb filters from BlueFors Parts No. 000105 are connected to the readout, microwave drive, and flux lines to minimize stray IR radiation coming from the SMA connectors. The measured voltage signals from the readout resonators are amplified by a series of low-noise amplifiers (e.g. Low Noise Factory LNF-LNC4-8G at $4\,$K, Narda-MITEQ LNA-40-04000800-12-15P at $300\,$K), and are upconverted with IQ mixers from Octave before the IQ signals are read by Quantum Machines' OPX+. Note that OPX+ with Octave can accommodate at most two qubits and two couplers in a fixed cabling session due to the limited number of analog-to-digital (ADC) input and receiver ports of OPX+, along with modulation and demodulation of IQ signals using Octave.

\begin{table}[htbp]
\caption{
Physical Parameters of the Transmon-Coupler System}
\begin{threeparttable}
\begin{ruledtabular}
\begin{tabular}{ccccc}
Transmon & C1 & Q1 & C2 & Q2\\
\colrule
\makecell{$f_{r}\,$\\(MHz)} & \multicolumn{2}{c}{$6073.7\,\pm\,0.2$} & \multicolumn{2}{c}{$5985.8\,\pm\,0.2$} \\
\makecell{$^{1}\kappa/2\pi$\\(kHz)} & \multicolumn{2}{c}{$108.8$ (\SI{91.9}{\micro s})} & \multicolumn{2}{c}{$273.4$ (\SI{36.6}{\micro s})} \\
\makecell{$f_{01,\text{max}}$\\(MHz)} & \makecell{8044.6 \\ $\pm\,0.2$} & \makecell{4768.6 \\ $\pm\,0.5$} & \makecell{7909 \\ $\pm\,2$} & \makecell{5081.9 \\ $\pm\,0.5$} \\
\makecell{$g_{qr}/2\pi\,$\\(MHz)} & $4\,\pm\,1$ & $94\,\pm\,2$ & $1.5\,\pm\,0.2$ & $95\,\pm\,2$ \\
\makecell{$E_{C}/h$\\(MHz)} & \makecell{120 \\ $\pm\,3$}  & \makecell{206 \\ $\pm\,3$} & \makecell{127 \\ $\pm\,3$} & \makecell{208 \\ $\pm\,3$} \\
\makecell{$T_{1}(\Phi=0)\,(\mu \textrm{s})$} & $4.8\,\pm\,0.4$  & $11.1\,\pm\,0.5$ & $4.9\,\pm\,0.9$ & $6.0\,\pm\,0.6$ \\
\makecell{$T_{2}(\Phi=0)\,(\mu \textrm{s})$} & $7\,\pm\,2$  & $4.8\,\pm\,0.4$ & $4.0\,\pm\,0.6$ & $8.0\,\pm\,0.6$ \\
\makecell{$^{2}T_{2}^{*}\,(\mu \textrm{s})$} & $2\,\pm\,2$ & $2.4\,\pm\,0.2$ & $3.3\,\pm\,0.4$ & $2.1\,\pm\,0.1$ \\
$^{3}F_a (\%)$ & 83.9 & 90.1 & 72.4 & 91.6 \\
\end{tabular}
\begin{tablenotes}
      \small
      \item $^{1}$ The values in parenthesis refer to $2\pi\times10\kappa^{-1}$
      \item $^{2}$ Ramsey $T_{2}$ time
      \item $^{3}$ $F_{a}$ refers to readout assignment fidelity.
    \end{tablenotes}
\end{ruledtabular}
\end{threeparttable}
\label{tab:dev_params}
\end{table}

Our single-qubit (1Q) gate tune-up follows standard calibration protocols for flux-tunable transmon qubits~\cite{Zchen_2018} to characterize the device parameters. Both microwave and flux pulses are generated using the RF and arbitrary waveform generator ports of the Quantum Machines OPX+ pulse processing unit with Octave. A Gaussian pulse with gate duration of \SI{40}{\nano s} is used to excite the qubit to its $|1\rangle$ and obtain the relaxation time $T_{1}$, and dephasing time $T_{2}^{R}$ by the Rabi and Ramsey pulse sequences, respectively. Readout assignment fidelities are optimized by single-shot measurements with integration times of \SI{0.4}{\micro s} for qubits and \SI{4}{\micro s} for couplers. 1Q gate fidelities of 99.6$\,\%$ for Q1 and Q2 were obtained with the microwave drive with derivative removal by the adiabatic gate (DRAG) sequence. A rectangular flux pulse with \SI{100}{\nano s} gate duration is used to tune the qubit frequency. Both microwave and flux pulses were used to perform two-tone spectroscopy, as shown in Figs. \textcolor{blue}{2} and \textcolor{blue}{3} of the main text. Other relevant parameters of the devices that are not discussed in the main text are listed in Table \ref{tab:dev_params}.   

\section{\label{sec:meas_methods}Calibration between Flux and Qubit-Coupler Interactions}

\subsection{\label{sec:sup_f01_V_conv} Mapping between Flux Bias Voltage and Transmon Idle Frequency}

The frequency of a flux-tunable qubit (or coupler), which has a SQUID loop as a tunable element, is governed by its Josephson energy \( E_J \), which depends on the external magnetic flux \( \Phi \). In the transmon limit ($E_J\gg E_{C}$), the $0\rightarrow 1$ transition frequency of the flux-tunable qubits and couplers \( f_{01} \) can be written as\cite{Koch2007}  
\begin{equation}
    f_{01}(\Phi) = \frac{\sqrt{8 E_C E_J(\Phi)} - E_C}{h},
    \label{eqn:f01_xmon1}
\end{equation}
with
\begin{equation}
    E_\text{J}(\Phi) \rightarrow E_{\text{J,max}} \textrm{cos}\left(\frac{\pi\Phi}{\Phi_0}\right) \sqrt{1+d^{2}\mathrm{tan}^{2}\frac{\pi\Phi}{\Phi_0}},
    \label{eqn:f01_xmon2}
\end{equation}
where \(E_{J1,\max}=E_{J1}+E_{J2}\) is the sum of energies of two junctions in a SQUID loop, \(d=(E_{J2}-E_{J1})/E_{J,\max}\) is the junction asymmetry, and \( E_C \) is the charging energy, which has opposite sign of the anharmonicity. 

The applied flux $\Phi$ that threads the $j^{th}$ SQUID loop is generated when the $i^{th}$ Z-line, which has mutual inductance with the SQUID loop, has current flowing through it. The applied current is proportional to the applied voltage $V_{\text{i}}$ across a line impedance of $50\,\Omega$. The voltage range used to tune $E_{J}$ allows the observation of one oscillating period of the resonator frequency in the dispersive readout without exceeding the critical current of the flux lines. 

Given this intuition, \eref{eqn:f01_xmon1} and \eref{eqn:f01_xmon2} can be rewritten as \cite{Barrett2023}
\begin{equation}
\begin{split}
f_{01}(V_i)&=\left(f_{\text{01,$\max$}}+\frac{E_C}{h}\right) \times \\  &\quad\sqrt[\leftroot{-2}\uproot{2}4]{d^2 + \left[ 1-d^{2} \right] \textrm{cos}^{2}\left(A_c \left[ V_i-V_{\text{ofs,}i}\right] \right)} \\ &\quad- \frac{E_C}{h},
\end{split}
\label{eqn:sf01_V}
\end{equation}
where $f_{01,\max}=(\sqrt{8E_{J,\max}E_{C}}-E_{C})/h$ is the frequency of the upper sweet spot, $-E_{C}/h$ is the anharmonicity, $A_{c}$ is a conversion factor related to the mutual inductance between the $i^{th}$ SQUID loop and the $i^{th}$ flux line, and $V_{\text{ofs,i}}$ is the offset voltage induced by remnant flux. Note that $A_{c}$ and $V_{\text{ofs,i}}$ are related to the normalized flux of the transmon by 
\begin{equation}
A_{c}(V_{i}-V_{\text{ofs,i}})=\frac{\pi\Phi}{\Phi_0}.
\label{eqn:norm_flux}
\end{equation}
We used the nominally designed junctions asymmetry noted in Sec.~\ref{sec:sup_fab} as the range of flux bias voltage used in Figs. \textcolor{blue}{2}(c-f) are not enough to extract $d$ using \eref{eqn:sf01_V}. Increasing the flux bias range to see the lower sweet spot frequency for the qubits would help extract $d_\text{Q}$ at the expense of additional measurement time and increased pixel resolution. This idea is not worthwhile for the tunable couplers as \eref{eqn:sf01_V} is only valid in certain flux bias regions where the coupler operates in the transmon regime (i.e., $E_{J,\text{C}}/E_{C,\text{C}}\geq30$). $E_{C}$ is determined by the two-photon excitation experiment. \eref{eqn:sf01_V} is used to fit the peak frequencies obtained from the qubit and coupler spectra of Figs. \textcolor{blue}{2}(c-f) in the main text, with $f_{\text{01,max}}$, $V_{\text{ofs,}i}$ and $A_{c}$ as fitting variables. \eref{eqn:norm_flux} is then used to calibrate the horizontal axis in Figs. \textcolor{blue}{2}(c-f) according to its normalized flux response. 

\eref{eqn:sf01_V} and \eref{eqn:norm_flux} show how junction asymmetry $d$ controls the tunable frequency range of the transmon according to the normalized flux quanta. Our couplers, which are transmons with symmetric junctions, have a gigahertz frequency in the tuning range, enabling coupler-mediated CZ coupling strengths in the tens of megahertz, and direct measurement of the coupling strength between the readout resonator and the coupler. Our qubits, which are transmons with asymmetric junctions $d_{Q}$, have limited frequency tuning within 3-5$\,$GHz, which is sufficient for 2Q gate operations, while minimizing qubit dephasing from flux noise~\cite{Braumuller_2020}, which scales with $df_\text{01,Q}/d\Phi\propto(1-d_{Q}^2)/d_{Q}$.

As two-qubit gate operations require accurate idling of the qubit and coupler frequencies away from the upper sweet spot, it is essential to determine the precise bias voltage. Solving the bias voltage as a function of $f_{01}$ from \eref{eqn:sf01_V}, the idle flux bias can be written as
\begin{equation}
\begin{split}
V_{\text{idle}}(f_{01})&= V_{\text{ofs},i} \pm \frac{1}{2A_{c}} \times \\ &\quad\textrm{arccos} \left( \frac{2}{1-d^2} \left[ \frac{f_{01} + E_C/h}{f_{01,\max} + E_C/h}\right]^{4} - \frac{1+d^2}{1-d^2} \right).
\end{split}
\label{eqn:sV_f_idle}
\end{equation}
\eref{eqn:sV_f_idle} helps determine idle voltages through an idle qubit and coupler frequencies that enable 2Q gates like the iSWAP gate and the conditional phase (CZ) gate.

\subsection{\label{sec:sup_flux_methods_1} Ramsey-based Protocol}

\begin{figure*}[t]
    \centering
    \includegraphics[width=17.2cm]{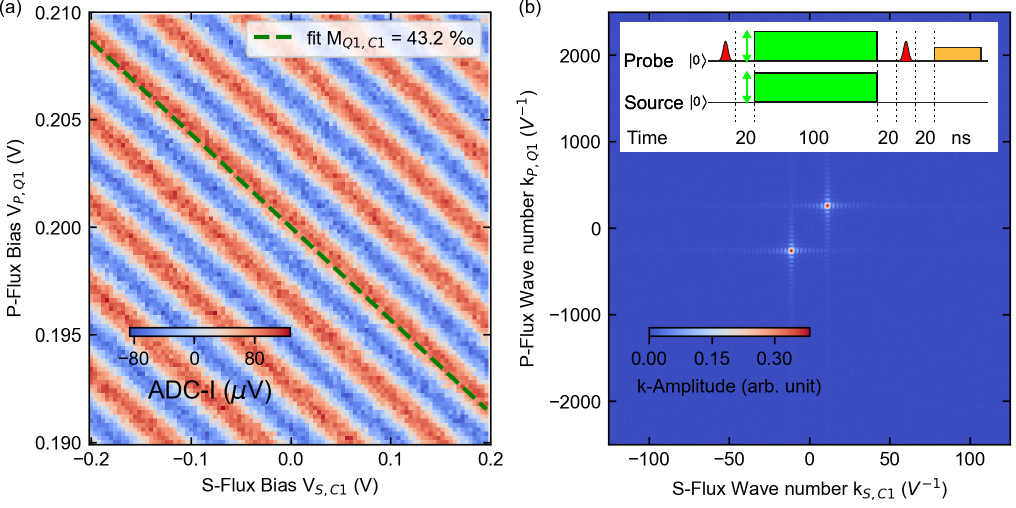}
    \caption{Characterizing magnetic flux crosstalk by resolving Ramsey Oscillations (a). By performing the 2D-FFT of the Ramsey fringes as shown in (b), we determine the perpendicular crosstalk. Inset shows the pulse schedule used to detect the flux crosstalk via Ramsey interferometry. By obtaining an orthogonal slope from the two points, we measure the flux crosstalk, which translates to a green-dashed line in (a).}
    \label{sfig:ramsey}
\end{figure*}

Figs.~\textcolor{blue}{3}(a) of the main text and the inset of \fref{sfig:ramsey}(b) show the MZLC and Ramsey-based pulse sequences, respectively. Both are used to measure flux crosstalk. In both methods, voltages are applied simultaneously to the probe Z-line and a source Z-line to differentiate flux crosstalk from other sources of nonlinear frequency shifts. The probe voltage is adjusted to resolve the unwanted frequency shift induced by the source voltage on the probe element. In the absence of flux crosstalk, there should be no frequency dependence of the probe element on the source voltages. Both source and probe voltages are converted into flux quanta to intuitively link the experiment to the transmon picture.

The two approaches differ only in the frequency extraction method. MZLC utilizes two-tone spectroscopy overlapping with probe and source voltages, while the Ramsey-based approach employs Ramsey interferometry, which outputs oscillating patterns shown in \fref{sfig:ramsey}(a). MZLC has good contrast for low readout SNR measurements and is robust against decoherence. The drawback of this method lies in its inhomogeneous broadening near the sweet-spot frequency, which will be discussed in detail in the latter subsections. 

Meanwhile, the Ramsey-based method~\cite{Neill2018,Sung_2021,Shi_2023_top} has increased sensitivity to frequency detuning induced by different probe bias ranges, leading to more precise flux crosstalk measurement, as shown from \fref{sfig:ramsey}(b). This measurement is limited by the qubit dephasing time. For the two-tone approach, both qubits are susceptible to dephasing due to an excess $|1\rangle$ population due to incomplete thermalization of the qubit and spurious excitation from the rectangular readout pulse. 

\subsection{\label{sec:sup_flux_methods_2} Flux Crosstalk Compensation Scheme}

The procedure for flux crosstalk cancellation has three steps. First, we measure the transition frequencies of all tunable qubits and couplers. Second, we quantify the crosstalk between a source Z-line and nearby detector elements. Finally, we construct a crosstalk cancellation matrix that enables effective multi-Z-line control.

Flux crosstalk arises from mutual inductive coupling between the flux lines (Z-lines) and the superconducting loops of nearby qubits or couplers. The mutual inductance matrix $L$ represents these sets of inductances and maps the vector of the Z-line current, represented as a matrix $I_Z$, to the vector of induced fluxes as follows:
\begin{equation}
    {\Phi} = L \cdot I_Z.
\end{equation}
Current is generated from the applied voltage $V_{Z}$ across a $50\,\Omega$ impedance of the Z-line. Generating all currents in the subsystem makes up an applied voltage matrix $V_{Z}$, we define an effective mutual inductance matrix $M$ accounting for the resistance and line response such that:
\begin{equation}
    {\Phi} = \mathbf{M} \cdot \mathbf{{V}_Z}.
\end{equation}
Since our goal is to cancel the undesired flux threading a detector SQUID loop, the absolute scaling in $M$ is irrelevant. We then normalize each row of $M$ by its diagonal entry $M_{ii}$, resulting a dimensionless crosstalk matrix $X$:
\begin{equation}
    V_\text{eff} = X \cdot V_Z.
    \label{eqn:X}
\end{equation}
We are concerned about the voltage change $\Delta V_{Z_\text{i}}$ required on the Z-line $i$ to compensate for the flux induced by a voltage change $\Delta V_{Z_\text{k}}$ on the Z-line $k$. We expect a vanishing net flux between flux lines:
\begin{equation}
    \Phi_i = M_{\text{ii}} \Delta V_{Z_{\text{i}}} + M_{\text{ik}} \Delta V_{Z_\text{k}} = 0.
    \label{eqn:compense}
\end{equation}
The solution for \eref{eqn:compense} defines an element in the flux crosstalk matrix $X$:
\begin{equation}
    X_\text{ik} = \frac{M_{\text{ik}}}{M_{\text{ii}}} = -\frac{\Delta V_{Z_\text{i}}}{\Delta V_{Z_\text{k}}}.
\end{equation}

\begin{figure*}[htbp]
    \centering          \includegraphics[width=1\textwidth]{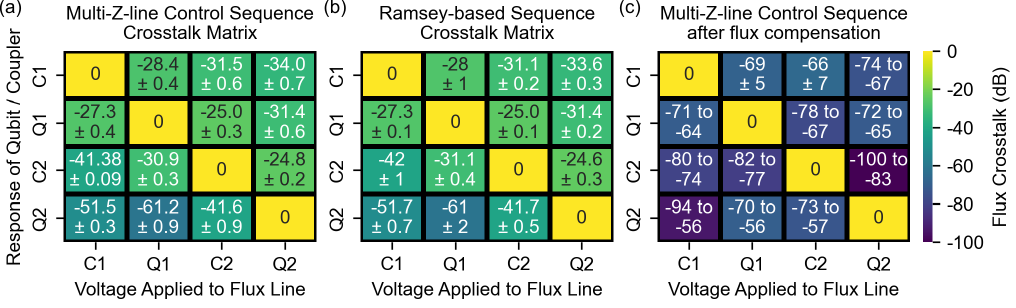}
    \caption{
    Flux crosstalk matrices X in the main-text but now expressed in dB. In (c), some elements have a lower bound of $-\infty$, due to the nominal values having lower magnitude than its standard deviation. Hence, we simply report those elements in range of dB magnitudes.}
    \label{fig:table}
\end{figure*}

In summary, all diagonal elements of $X$ are equal to 1 ($X_{\text{ii}}=1$), and the off-diagonal elements $X_{\text{ik},\text{i}\neq \text{k}}$, quantify the fraction of compensation voltage required on the Z-line $i$ to cancel the flux induced by the Z-line $j$. This formulation provides a simple and intuitive picture of flux crosstalk, directly in terms of the applied voltages. $X$ is characterized by both characterization protocols as shown in Fig. \textcolor{blue}{4}(a,b) of the main text with its absolute magnitude represented in \fref{fig:table}(a,b).

To eliminate flux crosstalk, we compute the inverse of this matrix, \( X^{-1} \), and use it to define a \textit{cancellation matrix} that maps the target bias voltages back to physical control voltages:
\begin{equation}
    V_Z = X^{-1} \cdot V_{\text{target}}.
\end{equation}
By applying these compensated voltages to the Z-lines, each tunable element can be independently biased, ensuring that off-diagonal flux contributions are suppressed. The effectiveness of the cancellation is verified by remeasuring the crosstalk matrix after compensation and confirming that it approximates the identity matrix as shown in Fig. \textcolor{blue}{4}(c) in the main text, and its absolute magnitude in \fref{fig:table}(c).

\subsection{\label{sec:sup_geff_Vc} Mapping between Coupler-Mediated CZ Coupling Strength and Flux-Compensated Flux Bias}

The CZ coupling strength in a transmon coupler architecture can then be written as~\cite{Sete_2021} 
\begin{equation}
g_{\text{eff}} = g_{02}= \sqrt{2}\left(g_{12} - \frac{g_{1c}g_{2c}}{2}B_{02} \right),
\label{eqn:sg_cz1}
\end{equation}
where $g_{12}$ is the direct coupling between the qubits Q1 and Q2, $g_{1c}$ is the coupling strength between the qubit Q1 and the coupler C2, $g_{2c}$ is the coupling strength between Q2 and the coupler C2, and~\cite{Sete_2021}
\begin{equation}
B_{02} = \frac{1}{\Delta_1} + \frac{1}{\Delta_2-E_{C,\text{Q2}}/\hbar} + \frac{1}{\sum_1} + \frac{1}{\sum_2 + E_{C,\text{Q2}}/\hbar}.
\label{eqn:sg_cz2}
\end{equation}
In \eref{eqn:sg_cz2}, $\Delta_{j}=\omega_c-\omega_{j}$ where $\omega_c=2\pi f_{01,c}$ is the coupler frequency and $\omega_j=2\pi f_{01,j}$ is the frequencies of the $j^{th}$ qubit. $\sum_j=\omega_{c}+\omega_{j}$ and $-E_{C,\text{Q2}}/\hbar$ is the anharmonicity of Q2. The first term in \eref{eqn:sg_cz2} represents the detuning between $\omega_{c}$ and idle $\omega_{01}$ of Q1 while the second term refers to the detuning between the coupler and the idle $\omega_{21}$ frequency of Q1. The third and fourth terms refer to the sum of the coupler frequency and the idle frequencies for the first and second qubits for CZ operation. 

Resonance points where \eref{eqn:sg_cz1} 1) approaches infinity and 2) becomes zero can be identified when the frequency detuning between $|11\rangle$ and $|02\rangle$ states $\delta_{21}=\omega_2 -(\omega_1+E_{C,\text{Q2}})=0$. This condition modifies \eref{eqn:sg_cz2} to
\begin{equation}
B_{02} = 4\left(\frac{\omega_c-E_{C,\text{Q2}}/\hbar}{\left[\omega_c-(E_{C,\text{Q2}}/\hbar)\right]^2-\omega_2^2}\right).
\label{eqn:sg_cz3}
\end{equation}
\eref{eqn:sg_cz3} blows up to infinity at $\omega_{c}=\omega_2+E_{C,\text{Q2}}/\hbar$, which is considered a CZ resonant point. Moreover, \eref{eqn:sg_cz3} leads to an approximate model of the coupler frequency where the CZ interaction is turned off. By evaluating \eref{eqn:sg_cz3} in \eref{eqn:sg_cz1} when $g_\text{eff}=0$, enforcing the coupler to operate in the transmon regime with frequencies $\omega_c>0$, and operating in the regime where $\omega_2\gg-E_{C,\text{Q2}}/\hbar>g_{1c},g_{2c},g_{12}$, and assuming a first-order Taylor series expansion of $\sqrt{(g_{1c}g_{2c}/g_{12})^2+\omega_2^2}$, the coupler frequency at the CZ off-point can be written as
\begin{equation}
\omega_c \approx \omega_2 + \frac{E_{C,\text{Q2}}}{\hbar}+ 2\left(\frac{g_{1c}g_{2c}}{g_{12}}\right)+\frac{1}{2}\left(\frac{g_{1c}g_{2c}}{g_{12}}\right)^2\left(\frac{1}{\omega_2}\right).
\label{eqn:swc_off}
\end{equation}
Both \eref{eqn:sg_cz3} and \eref{eqn:swc_off} provide analytical models that describe the relevant frequency bounds of the coupler for CZ interaction in the case that the coupler sweet spot frequency is higher than the sweet spot frequency of the qubits. Note that these were previously qualitatively determined for an iSWAP case~\cite{Liang_2023_swap}.

We then characterize the coupler-mediated coupling strength $g_\text{eff}$ between states $\ket{11}$ and $\ket{02}$. Here, the first and second digits of the state vectors refer to the single-qubit state of Q1 and Q2, respectively. After preparing a $\ket{11}$ state, we apply a rectangular pulse with pulse duration $\text{T}=100\,$ns for both qubits and couplers. Assuming that the detuning $\delta_{21}$ between the $\ket{11}$ and $\ket{02}$ states has observable Rabi oscillations, the population of the $\ket{11}$ for two coupled qubits can be modelled as~\cite{Sung_2021,hellings_2025,Krizan_2025} 
\begin{equation}
P_{\ket{11}\rightarrow\ket{02}}(\delta_{21},g_{\mathrm{eff}},t)=\frac{2g_{\mathrm{eff}}^{2} \left[1+\textrm{cos} \left(D \times t\Omega(g_{\mathrm{eff}},\delta_{21})\right) \right]}{\Omega^{2}(g_{\mathrm{eff}},\delta_{21})},
\label{eqn:sP_11}
\end{equation}
where $t$ is the duration between diabatic pulses,  
\begin{equation}
\Omega^{2}(g_{\mathrm{eff}},\delta_{21})=4g_{\text{eff}}^{2}+\delta_{21}^{2},
\label{eqn:somega}
\end{equation}
and $D$ represents the reduced factor accounting the imperfect diabatic pulse shape experienced by the coupler. We model $D$ as a ratio of a diabatic pulse with an effective rising pulse time $t_\text{eff}$ and its corresponding amplitude $A_\text{eff}$ to that of the ideal pulse shape area with XY amplitude $\Omega=A$ in angular frequency units and pulse time duration $\text{T}$:
\begin{equation}
D=\frac{\int_{0}^{\text{T}} A+A_\text{eff}\times\textrm{exp}(-t / t_\text{eff}) \, dt }{A\text{T}}.
\label{eqn:sD_ratio}
\end{equation}

\eref{eqn:sP_11} simulates the CZ SWAP experiment at constant $\delta_{21}$, and the CZ leakage at constant $\text{t}$ as shown in Fig. 5(d) of the main text. Fig. 5(d) is in excellent agreement with Fig. 5(c) in the main text when we evaluate \eref{eqn:sP_11} with the following qubit, coupler and $D$ parameters extracted in the experiment: $\omega_\text{1,idle}/2\pi=4720\,$MHz, $\omega_\text{2,idle}/2\pi=4931\,$MHz, and $E_{C, \text{Q2}}/2\pi\hbar=202\,$MHz. Also, the coupler C1 is detuned at frequencies $\omega_\text{c1, idle}/2\pi=7735\,$MHz when performing the CZ SWAP experiment. Here, $D=0.16$ is a fitting parameter that is related to the number of Rabi oscillation cycles observed on the CZ SWAP map.

For \eref{eqn:sD_ratio} to achieve a value $D=0.16$ that makes Fig. 5(d) in excellent agreement with Fig. 5(c) of the main text, we set $A_\text{eff}=-A$ and $t_{\text{eff}}=278\,$ns. This value of the time constant, though extracted empirically, accounts for both short and long time constants that were not corrected even after flux compensation. These details will be thoroughly discussed in the next subsection. 

Proper mapping between the idle voltage of the coupler $V_{c}$ and the coupling strength $g_{\text{eff}}$ is a key engineering trick to control the interaction strength between two qubits in a CZ gate operation. Note that the dependence between $g_{\text{eff}}$ and $\delta_{21}$ can be traced on the idling of the second qubit in its $|1\rangle\rightarrow|2\rangle$ frequency
\begin{equation}
\omega_{2} = \delta_{21} + \omega_{1} + E_{C,\text{Q2}}/\hbar.  
\label{eqn:w_q2_cz}
\end{equation}

As $\omega_{2}$ are present in the second, third and fourth term in \eref{eqn:sg_cz2}, solving for $\omega_{1}$ as functions of $g_{\mathrm{eff}}$ and $\delta_{21}$ becomes cumbersome, and requires numerical routines. Once $\omega_{1}$ is solved, we use \eref{eqn:sf01_V} to express the coupler voltage as functions of $g_{\mathrm{eff}}$ and $\delta_{21}$. Establishing its analytical form for intuition requires 1) precise control of parameters such as $\delta_{21}=0$ so that $\omega_{c}-\omega = \Delta_1 = \Delta_2-E_{C,\text{Q2}}/\hbar$ and 2) neglecting the third and fourth terms in \eref{eqn:sg_cz2}. These are valid for $\omega_c\gtrsim\omega$.

The simplifications made above rewrite \eref{eqn:sg_cz1} and \eref{eqn:sg_cz2} as 
\begin{equation}
g_{\text{eff}} = \sqrt{2}\left(g_{12} - \frac{2g_{1c}g_{2c}}{\omega_c-\omega} \right).
\label{eqn:sg_cz_approx}
\end{equation}
By rearranging the terms to solve for $\omega_{c}=2\pi f_{01,c}$ and plugging in \eref{eqn:sf01_V} to solve for $V_{\text{idle},c}$, the idle flux bias of the coupler required to control the coupling strength in a CZ interaction can be written as
\begin{equation}
\begin{split}
V_{\text{idle},c}(g_{\text{eff}},\omega) &=V_{\text{ofs},c} \pm \frac{1}{2A_{c,cj}} \times \\  &\quad\textrm{arccos} \left(\frac{2}{1-d_c^2} \left[K(g_{\text{eff}},\omega)\right]^{4} - 1 \right),
\end{split}
\label{eqn:sVidle_gcz1}
\end{equation}
where
\begin{equation}
K(g_{\text{eff}},\omega)=\frac{\omega + \frac{2g_{1c}g_{2c}}{g_{12}-g_{\text{eff}}/\sqrt{2}}-E_{C,\text{C2}}/\hbar}{\omega_{01,c,\max}+E_{C,\text{C2}}/\hbar}.
\label{eqn:sVidle_gcz2}
\end{equation}

As \eref{eqn:sVidle_gcz1} and $\omega$ assume appropriate flux crosstalk compensation for precise idling and control of the coupler-assisted coupling strength as established in the main text, flux crosstalk compensation becomes a vital protocol in accurately setting not only the idle qubit frequencies in a two-qubit gate operation, but also in precise control of $g_{\mathrm{eff}}$.

\subsection{\label{sec:sup_flux_trans} Mapping the Rabi Angle in the Empirical Rabi model with the Flux Transients}

The effective exponential decay function that makes $0<D\leq1$, as noted in \eref{eqn:sD_ratio}, can be represented as the sum of multiple exponential decay time constants~\cite{TMLi2024,Sung_2021,hellings_2025} that distort the flux pulse in \eref{eqn:sD_ratio}: 
\begin{equation}
A_\text{eff}\times\textrm{exp}(-t / t_\text{eff})=\sum_{m=0}^{n} a_m \text{exp}({-t/t_m}),
\label{eqn:seff_flux_trans}
\end{equation}
where $a_m$ and $t_m$ are the individual amplitude and decay time of the flux transients that range from nanoseconds to hundreds in microsecond timescales~\cite{Sung_2021,TMLi2024,hellings_2025}. From \eref{eqn:seff_flux_trans}, $A_\text{eff}$ and $t_\text{eff}$ can be written in terms of the individual transient time and decay time. 
\begin{equation}
A_\text{eff}=\sum_{i=0}^{m} a_i
\label{eqn:seff_flux_trans_amp}
\end{equation}
\begin{equation}
t_\text{eff}=\left(\sum_{i=0}^{m}\ a_i \right) / \left(\sum_{i=0}^{m} \frac{a_i}{t_i}\right). \\
\label{eqn:seff_flux_trans_teff}
\end{equation}
Note that $A_\text{eff}$ and $t_\text{eff}$ in \eref{eqn:sD_ratio} and \eref{eqn:seff_flux_trans} are not meant to be used in mitigating these flux transients via flux pulse predistortion as done in other works~\cite{Sung_2021, Shi_2023_cal}. \eref{eqn:seff_flux_trans}, when considered in \eref{eqn:sD_ratio} and \eref{eqn:sP_11}, shows how the sum of these flux transients could possibly reduce the rate of $|11\rangle \leftrightarrow |02\rangle$ exchange.

Given the aforementioned parameters for a CZ SWAP experiment as discussed in \ref{sec:sup_geff_Vc}, setting $D=1$ would produce 17 Rabi oscillation cycles, which is larger than the 3 Rabi oscillation cycles shown in Figs. 5(c) and 5(d) of the main text. For a given flux pulse, a smaller integrated Rabi angle leads to 1) lower maximum $|02\rangle$ excitation probability and 2) reduced interaction time between $|11\rangle$ and $|02\rangle$, which affects the accumulated phase and the fidelity of the CZ gate. 

\subsection{\label{sec:sup_xtalk_vs_time} Dependence of flux crosstalk with varying pulse duration}

\begin{figure}[htbp]
    \centering          \includegraphics[width=8.6cm]{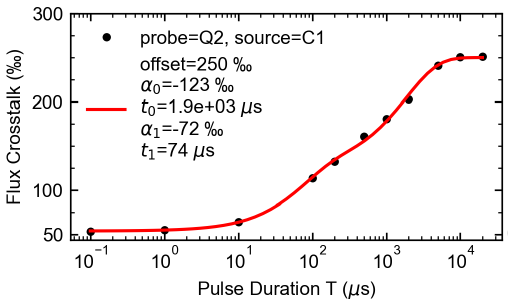}
    \caption{
    Flux crosstalk measurement on Q2 as the probe and C1 as the source by varying pulse duration $T$. The flux crosstalk was initially measured to be 50$\,$\textperthousand$\,\,$at pulse duration T=$100\,$ns and increases exponentially until it saturates to 250$\,$\textperthousand$\,\,$at T=\SI{20}{\milli s}. The red line refers to the best fit according to a multi-exponential model with a saturation offset of 250$\,\,$\textperthousand.}
    \label{fig:s_xtalk_vs_time}
\end{figure}

We have examined the pulse duration dependence of the crosstalk matrix element between the probe Q2 and the source C1 in another sample with the same layout. As shown in \fref{fig:s_xtalk_vs_time}, the crosstalk amplitude extracted gradually increases with longer flux pulse durations, from approximately 0.05$\,$\textperthousand $\,$at \SI{0.1}{\micro s} to 0.25$\,$\textperthousand$\,$at \SI{20}{\milli s}, and then saturates. The data points are then fitted with a constant offset added to the multi-exponential time constant function as expressed in \eref{eqn:seff_flux_trans}. 

We identified two long-time constants at tens of milliseconds and \SI{1.9}{\milli s}. Typical origins for timescales of tens of milliseconds belong to eddy currents created on the surface of the PCB board, which have characteristic L/R times~\cite{Foxen_2019}. Future investigation may be needed to resolve these time-constant origins, especially in the millisecond mark. By eliminating the source of the long millisecond time constant, and predistorting the flux pulse to remove the effects of these time constants before flux crosstalk cancellation, the value of $D$ could approach unity~\cite{Barends2014,Sung_2021,TMLi2024}. 

\section{\label{sec:sup_broaden} Flux-Induced Inhomogeneous Effects on Linewidth in MZLC}

\begin{figure}[htbp]
    \centering          \includegraphics[width=8.6cm]{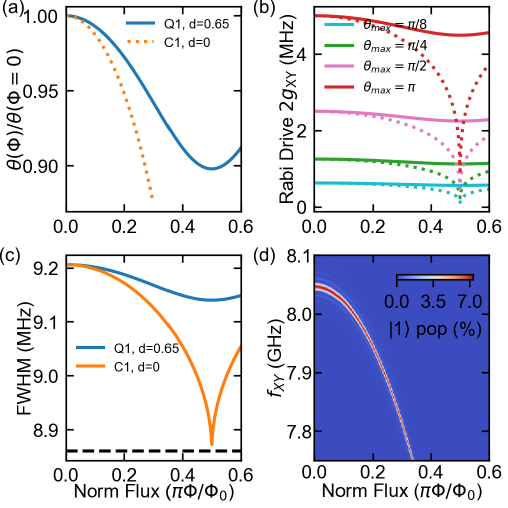}
    \caption{
Simulated effects of flux bias on dipole matrix element. (a) The normalized Rabi angles of Q1 (blue, solid lines) and C1 (orange, dotted line) are flux-modulated. The maximum Rabi angle is at $\Phi=0$. (b) Rabi XY drive frequency as a function of normalized flux bias. Different colors refer to the strength of the Rabi driving, controlled by the Rabi angle. The simulated Rabi frequency of Q1 and C1 is shown in solid and dotted lines, respectively. (c) Flux-dependent linewidth broadening of Q1 (blue) and C1 (orange). Black dashed lines, meanwhile, refer to a Fourier-limited linewidth of \SI{8.8}{\mega Hz}, set by the \SI{100}{\nano s} XY drive. We use $\theta_{max}=\pi/8$ for the simulation. (d) Simulated transmon spectra of C1 as a function of normalized flux and XY driving frequency. Linewidth broadening is significant at the highest transmon frequency. Refer to Sec.~\ref{sec:sup_fab} and Table~\ref{tab:dev_params} for the relevant parameters for Q1 and C1.}
    \label{fig:s_broaden}
\end{figure}

Any constant driving power in two-tone spectroscopy may not be truly uniform, as the transmon receives the microwave drive differently at a calibrated XY driving power. Suppose we apply an XY drive to the qubit and calibrate this power at the maximum qubit frequency, $f_{01}(\Phi=0)$. We then define the Rabi frequency between the qubit transition and the XY drive as~\cite{Sung_2022}
\begin{equation}
2g_\text{XY,i}=\beta\frac{V_\text{XY}(2e)\langle i|\hat n|i+1\rangle}{\hbar},
\label{eqn:s_rabi_xy1}
\end{equation}
where $\beta=C_\text{XY}/(C_\text{XY}+C_\text{xmon})$ is the participation ratio of the XY drive in the total circuit capacitance, which includes the transmon capacitance $C_\text{xmon}$, $V_\text{XY}$ is the XY drive voltage on the transmon, $e$ is the electron charge and $\langle i|\hat n|i+1\rangle$ is the dipole matrix element of the transmon between the $i^{th}$ and $()i+1)^{th}$ energy levels. In the limit of $E_j\gg E_C$, and evaluating the coupling strength $2g_\text{XY}$ of the XY drive, the dipole matrix element of the transmon can be expressed as 
\begin{align}
\label{eqn:s_rabi_xy2a}
g_\text{XY,i} &=\sqrt{i+1}g_\text{XY,0}=\sqrt{i+1}g_\text{XY} \\ 
\langle i|\hat n|i+1\rangle &\approx \sqrt{\frac{i+1}{2}} \left( \frac{E_j(\Phi)}{8E_{C}} \right)^{1/4}
\label{eqn:s_rabi_xy2b}
\end{align}
Since our concern is the population between the ground and excited states in pulsed two-tone spectroscopy, we evaluate $i=0$, and evaluate \eref{eqn:s_rabi_xy1} together with \eref{eqn:s_rabi_xy2a} and \eref{eqn:s_rabi_xy2b}. The Rabi XY drive therefore depends on the applied flux as follows:  

\begin{equation}
2g_\text{XY}(\Phi)\approx\frac{2\sqrt{2}\beta V_\text{XY}}{\hbar} \left( \frac{E_j(\Phi)}{8E_{C}} \right)^{1/4} 
\label{eqn:s_rabi_xy3}
\end{equation}

For a rectangular pulse, we write the Rabi frequency in terms of the Rabi angle $\theta$
\begin{equation}
2g_\text{XY}(\Phi)=\frac{\theta(\Phi)}{\text{T}},
\label{eqn:s_rabi_theta}
\end{equation}
where $\text{T}\ll T_{1},T_{2}^{r}$ corresponds to the pulsed two-tone spectroscopy regime. We calibrate the Rabi XY drive amplitude at the maximum transmon frequency $f_{01}(\Phi=0)$, where the maximum Rabi XY angle reads as
\begin{equation}
\theta_\text{max}=\frac{2\sqrt{2}\beta V_\text{XY}}{\hbar} \left( \frac{E_{\text{J,,max}}}{8E_{C}} \right)^{1/4}\text{T}. 
\label{eqn:s_rabi_theta_max}
\end{equation}
As detuning the transmon with $\Phi$ modulates the Josephson energy through the trigonometric components in \eref{eqn:f01_xmon2}, the Rabi XY frequency is written as:
\begin{equation}
\theta(\Phi)=\theta_\text{max}\left(\left| \text{cos}\left(\frac{\pi \Phi}{\Phi_0} \right) \right|\sqrt{1+d^2\text{tan}^2\left(\frac{\pi \Phi}{\Phi_0} \right)}\right)^{1/4}.
\label{eqn:s_rabi_theta_phi}
\end{equation}

The flux-dependent Rabi XY drive frequency now reads as
\begin{equation}
2g_\text{XY}(\Phi)=2g_\text{XY,max}\left(\left| \text{cos}\left(\frac{\pi \Phi}{\Phi_0} \right) \right|\sqrt{1+d^2\text{tan}^2\left(\frac{\pi \Phi}{\Phi_0} \right)}\right)^{1/4},
\label{eqn:s_flux_rabi}
\end{equation}
where $\Omega_\text{max}$ is related to $\theta_\text{max}$ by \eref{eqn:s_rabi_theta}.

In pulsed two-tone spectroscopy, the $|1\rangle$ population can be written similarly to \eref{eqn:sP_11} for a rectangular pulse,
\begin{equation}
P_{|0\rangle\rightarrow|1\rangle}(\delta_\text{XY},\Phi)=\frac{2g_\text{XY}^2(1+\text{cos}(\Omega_\text{XY}\text{T}))}{\Omega_\text{XY}^2 },
\label{eqn:sPe}
\end{equation}
where $\Omega_\text{XY}^2(\delta_\text{q,XY},\Phi)=4g_\text{XY}^2(\Phi)+\delta_\text{q,XY}^2(\Phi)$, with $\delta_\text{q,XY}$ is the detuning between the transmon frequency and the XY drive. Taking the Fourier transform of \eref{eqn:sPe} would yield a $\text{sinc}^2$ lineshape, limited by the Fourier-limited linewidth of 0.886/T for weak Rabi drives 2$g_\text{XY}\text{T}\ll1$~\cite{Taylor_1999}. At high amplitudes, small detuning, and neglecting the oscillation component in \eref{eqn:sPe}, the resulting lineshape becomes Lorentzian, limited by the Rabi drive frequency~\cite{Mihov_2024}. An effective Rabi spectroscopy linewidth can be written as~\cite{Taylor_1999,Mihov_2024}:
\begin{equation}
\text{FWHM}(\Phi)=\sqrt{\left(\frac{0.886}{\text{T}}\right)^2+4g_\text{XY}^2(\Phi)}.
\label{eqn:sFWHM_Rabi}
\end{equation}

\eref{eqn:sFWHM_Rabi} works only to single-shot spectroscopy. In the later section of this work, most of the qubit and MZLC spectra presented in the main text have undergone many averages to improve the signal-to-noise (SNR) of the qubits and couplers. Low-frequency drifts, such as qubit-frequency fluctuations caused by flux noise, are thereby averaged out. Dephasing of the qubit by Gaussian-distributed noise~\cite{Martinis_2003} still occurs, leading to inhomogeneous linewidth broadening~\cite{Blais_2021,Krantz_2019} in spectroscopy. 

The model for inhomogeneous broadening needs \eref{eqn:sPe} to include an exponential decay function, with the decay constant being the dephasing time $T_{2}^{*}$. However, having an estimate of the dephasing time while performing spectroscopic averaging is challenging. Empirically, convolving \eref{eqn:sPe} captures the observed broadening in experiments where the linewidths resemble a Gaussian-like lineshape. An approximate analytical estimate of the empirical linewidth can be obtained by adding a Gaussian-filtered linewidth term, $(2.335 \times \sigma_\text{drift})^2$ to the square of \eref{eqn:sFWHM_Rabi} and applying the square root.

In the following section, we discuss how the spectroscopic SNR, and curve fitting allow the frequency shift uncertainty to go below the measured spectroscopic linewidth. 

\section{\label{sec:sup_snr} Spectroscopic Signal-to-Noise Ratio and Frequency shift Uncertainty of MZLC}
\subsection{\label{sec:sup_specsnr} Spectroscopic and Readout SNR}

The readout fidelity of superconducting qubits suffers from at least three sources of errors: readout, relaxation, and separation. We focus on the separation error, which is related to the single-shot readout SNR by ~\cite{Sank_2014}: 
\begin{equation}
\epsilon_\text{sep}=\frac{1}{2}\text{erfc}\left( \sqrt{\frac{\text{SNR}_\text{read}}{2}}\right),
\label{eqn:s_seperr}
\end{equation}
where the $\text{SNR}_\text{read}$ is given by ~\cite{Sank2025}:
\begin{equation}
\sqrt{\text{SNR}_{\text{read}}}=\frac{\left| Z_{|0\rangle}-Z_{|1\rangle} \right|}
{\sqrt2\sigma_m}=\frac{\left| \Delta Z \right|}{\sqrt{2}\sigma_m}.
\label{eqn:s_snr1}
\end{equation}
The first equality originates from the exponential terms of the Gaussian-distributed IQ blobs; $Z_{|0\rangle}$ and $Z_{|1\rangle}$ are integrated voltages in the IQ plane that represent the $|0\rangle$ and $|1\rangle$ states, respectively. $\sigma_{m}$ is the standard deviation of the Gaussian-distributed IQ blobs in units of voltage. The numerator defines the separation between ground and excited states, while the denominator represents the effective diameter of the IQ blobs, which sets the readout noise level in Sec.~\ref{sec:meas_methods} if properly calibrated. 

Next, we consider an integrated single-shot spectrscopic SNR that accounts for the populations of the ground state $P_{|0\rangle}$ and excited state $P_{|1\rangle}$. The spectroscopic signal reads as 
\begin{align}
Z_{spec}&=P_{|0\rangle}Z_{|0\rangle}+P_{|1\rangle}Z_{|1\rangle} \\
&=(Z_{|0\rangle}-P_{|1\rangle}Z_{|0\rangle})+P_{|1\rangle}Z_{|1\rangle} \\
&=Z_{|0\rangle}-P_{|1\rangle}\Delta Z.
\label{eqn:s_snr2}
\end{align}
When the signal is subtracted by the background spectroscopic signal, with the same form as \eref{eqn:s_snr2} but including a background excited state population $P_{|1\rangle\text{,off}}$ due to unwanted thermal excitation, the spectroscopic difference now becomes
\begin{equation}
|\Delta Z_\text{spec}|=(P_{|1\rangle}-P_{|1\rangle\text{,off}})|\Delta Z|.
\label{eqn:s_snr3}
\end{equation}
Since both measurements experience the same noise for the output line, the single shot spectroscopic SNR is
\begin{equation}
\text{SNR}_\text{spec}^\text{single}=\frac{|\Delta Z_\text{spec}|}{\sqrt{2}\sigma_m}.
\label{eqn:s_snr4}
\end{equation}
Substituting \eref{eqn:s_snr1} and \eref{eqn:s_snr3} into \eref{eqn:s_snr4} gives the spectroscopic SNR as a function of $\text{SNR}_\text{read}$, 
\begin{equation}
\text{SNR}_\text{spec}^\text{single}=(P_{|1\rangle}-P_{|1\rangle\text{,off}})\sqrt{\text{SNR}_\text{read}}.
\label{eqn:s_snr5}
\end{equation}
When averaging over $n_\text{avg}$ repetitions per pixel in pulsed two-tone spectroscopy, the spectroscopic SNR scale with $\sqrt{n_\text{avg}}$. Assuming $P_{|1\rangle\text{,off}}\approx0$ and substituting $P_{|1\rangle}$ from \eref{eqn:s_snr4} into \eref{eqn:sPe}, the averaged spectroscopic SNR is
\begin{equation}
\text{SNR}_\text{avg}(\delta_\text{q,XY},\Phi)=P_{|0\rangle\rightarrow|1\rangle}(\delta_\text{q,XY},\Phi)\sqrt{n_\text{avg}\text{SNR}_\text{read}}.
\label{eqn:s_snr6}
\end{equation}

Based on \eref{eqn:s_snr6}, the average spectroscopic SNR depends on three metrics: 1) the readout SNR, 2) the number of averages per pixel, and 3) $|1\rangle$ population excited by our XY pulse. A coarse-optimized readout and a sufficiently XY pulse are generally able to characterize the flux crosstalk early in the superconducting qubit calibration stack, along with multiple averages. However, having a good calibrated $\pi_{01}$ and $\pi_{01}/2$ pulses, and a good readout SNR helps reduce the required time for flux crosstalk characterization, and avoids the inhomogeneous broadening described in Sec.~\ref{sec:sup_broaden} from appearing in the pulsed two-tone spectroscopy maps.     

\subsection{\label{sec:sup_peakthres} Frequency uncertainty from Gaussian peak thresholding and curve fitting}

Peak thresholding in a 2D map obtained from pulsed two-tone experiments rapidly extracts transmon frequencies while suppressing the frequency uncertainty below the observed spectroscopic linewidth. However, uncertainties in the extracted frequencies are rarely understood intuitively in qubit experiments. Knowing how frequency precision narrows through peak thresholding and curve fits allows one to assess the trade-off between precision over calibration time when extracting transmon frequency shifts and precisions across varying ranges of biases.

In pulsed two-tone experiments, Gaussian (or Lorentzian) lineshapes best characterize the transmon frequency and spectral linewidth depending on the XY drive power and pulse duration. In the this work, we use a Gaussian lineshape with amplitude $A$, center frequency $f_0$, and standard deviation $\sigma$ on low XY drive power in pulsed spectroscopy. The choice corresponds to weak pulses lying between the Gaussian linewidth broadening observed in continuous wave power-broadening experiments experiments~\cite{Schuster_2005,Blais_2021} and the Lorentzian lineshape observed in Rabi-limited drive~\cite{Mihov_2024}. The Gaussian lineshape has the form 
\begin{equation}
g(f)=\text{exp}\left( -\frac{(f-f_0)^2}{2\sigma^2}\right),
\label{eqn:s_peak_0}
\end{equation}
where the Gaussian linewidth is written as 
\begin{equation}
\text{FWHM}=2\sqrt{2\text{ln}2}\sigma\approx2.355\sigma.
\label{eqn:s_FWHM_g}
\end{equation}
Here, the experimental spectroscopic SNR is defined as
\begin{equation}
\text{SNR}_\text{avg}=C/\sigma_{n},
\label{eqn:s_peak_1}
\end{equation}
where $\sigma_{n}$ is the limiting noise level in the measurement. A local threshold is set high enough value to filter peaks from the noisy baseline. The threshold is defined as $sC$ so $0<s<1$ and the crossing occurs at $f=f_0\,\pm\,\delta_s$ where $\delta_s$ are two frequencies where $f_0$ is at the midpoint. The prescribed condition leads to  
\begin{align}
\text{exp}\left( -\frac{\delta_s^2}{2\sigma^2}\right)=s,
\label{eqn:s_peak_2}
\end{align}
and thus:
\begin{align}
\delta_s= \sigma\sqrt{-2\text{ln}s}.
\label{eqn:s_peak_3}
\end{align}
Differentiating \eref{eqn:s_peak_0} with respect to $f$ and evaluating at the crossing $f=f_0\,\pm\,\delta_s$, the magnitude of the slope is 
\begin{align}
\left| \frac{dg}{df} \right|_{f_0\pm\delta_s}=C\frac{\delta_s}{\sigma^2}\text{exp}\left( -\frac{\delta_s^2}{2\sigma^2} \right).
\label{eqn:s_peak_4}
\end{align}
Substituting $\delta_s$ with \eref{eqn:s_peak_3} and the exponential with \eref{eqn:s_peak_2}, the local threshold slope is
\begin{align}
\left| \frac{dg}{df} \right|_\text{thr}=\frac{C}{\sigma}s\sqrt{-2\text{ln}s}.
\label{eqn:s_peak_5}
\end{align}

A small vertical noise $\delta y$ could cause a frequency shift $\delta f\approx\delta y/|dg/df|$. Therefore, the RMS standard deviation of a single crossing is
\begin{equation}
\sigma_{f_0,\text{cross}}=\frac{\sigma_n}{|dg/df|_\text{thr}}=\frac{\sigma_n\sigma}{C}\cdot\frac{1}{s\sqrt{-2\text{ln}s}}.
\label{eqn:sigma_f_cr}
\end{equation}
Since the center frequency is roughly at the midpoint between the two crossings, and assuming they are independent and have equal variance, the variance of the midpoint is half the variance of a single crossing. The root-mean-square (RMS) uncertainty of the center frequency is
\begin{align}
\sigma_{f_0,\text{s}}&=\frac{\sigma_n{\sigma}}{\sqrt{2}C}\cdot\frac{1}{s\sqrt{-2\text{ln}s}} \\
&=\frac{1}{4\sqrt{\text{ln}2}}\cdot\frac{\text{FWHM}}{\text{SNR}_\text{avg}}\cdot\frac{1}{s\sqrt{-\text{ln}s}}.
\label{eqn:s_sigma_thres0}
\end{align}
For the special case where the threshold is set just above the half-maximum threshold $s=0.5$
\begin{equation}
\sigma_{f_0,\text{s=0.5}} \approx 0.7213\frac{\text{FWHM}}{\text{SNR}_\text{avg}}.
\label{eqn:s_sigma_thres1}
\end{equation}

Since the experimental spectral SNR is obtained as the ratio between the amplitude obtained from the Gaussian peak and its noisy baseline, the linewidth derived from thresholding is several times smaller than the Fourier-limited linewidth of \SI{8.8}{\mega Hz} shown in \sfref{fig:s_broaden}(c), even in the presence of inhomogeneous broadening. 

\subsection{\label{sec:sup_curvefit} Frequency shift uncertainty by curve fit}

The transmon frequency precision improves when using curve fitting. Taking the Gaussian lineshape in \eref{eqn:s_peak_0} as an example, the precision can be enhanced by obtaining many independent samples across the peak. A natural definition for this sampling density is
\begin{equation}
n_\text{FWHM}=\frac{\text{FWHM}}{\delta_f},
\label{eqn:s_crlb0}
\end{equation}
where $\delta_{f}$ is the frequency spacing between sampling points. Assuming a uniform noise process in a 1D slice of the pulsed two-tone spectroscopy data shown in Figs.2(c) to 2(f), the lower bound of the standard deviation for \eref{eqn:s_peak_0} reads as~\cite{Hagen_2007} 
\begin{equation}
\sigma_{f_0,\text{fit}}^2\approx\frac{2\sigma^2_n \sigma}{C^2\delta_\text{px}\sqrt{\pi}},
\label{eqn:s_crlb1}
\end{equation}
where $\delta_\text{px}$ is the pixel width. It is useful to relate this pixel width to a uniform frequency spacing $\delta_f =u \delta_\text{px}^{-1}$, with $u$ as a scaling or conversion factor. Setting $u=1$ for convenience, the RMS uncertainty of the fitted frequency becomes 
\begin{equation}
\sigma_{f_0,\text{fit}}\approx\frac{\sigma_n}{C}\sqrt{\frac{2 \sigma\delta_\text{f}}{\sqrt{\pi}}}=\frac{1}{\text{SNR}_\text{avg}}\sqrt{\frac{2 \sigma\delta_\text{f}}{\sqrt{\pi}}}.
\label{eqn:s_crlb2}
\end{equation}
Substituting $\delta_f$ in \eref{eqn:s_crlb2} from \eref{eqn:s_crlb0} and $\sigma$ from \eref{eqn:s_FWHM_g} leads to a redefinition of the frequency standard deviation
\begin{equation}
\sigma_{f_0,\text{fit}}=G\frac{\text{FWHM}}{\text{SNR}_\text{avg}\sqrt{n_\text{FWHM}}},
\label{eqn:s_crlb3}
\end{equation}
where $G=0.69264$ for a pure Gaussian lineshape. 

The intuition behind \eref{eqn:s_crlb3} becomes clear: if an incorrect model is fitted to a spectral lineshape, the coefficient $G$ changes accordingly. The ratio between \eref{eqn:s_crlb3} and \eref{eqn:s_sigma_thres1} sampling points within the $n_\text{FWHM}^{-1/2}$. For many sampling points within the FWHM of the transmon lineshape, the frequency uncertainty can be reduced by several orders of magnitude through proper curve fitting, compared to extracting the frequency uncertainty directly from the observed spectral linewidth. In general, any other curve-fitting model under a uniform noise process follows a similar scaling of the frequency standard deviation, decreasing as $n_\text{FWHM}^{-1/2}$. 

\section{\label{sec:sup_1Q_fidelity} Effect of Flux Crosstalk Compensation on Single-Qubit Gate Fidelities}

Flux crosstalk may play an important role in maintaining the fidelity of simultaneous 1Q gate on a multiple qubits QPU. Note that flux crosstalk compensation does not improve 1Q gate fidelity of isolated flux-tunable elements that are separated by ground planes and fixed-frequency qubits. Such methods become relevant for flux-sensitive elements affected by flux crosstalk, such as coupled multiqubit systems that have dedicated or shared flux lines that simultaneously bias at least two elements while other qubits operate 1Q gates. When transmons are flux-biased with dedicated flux lines to characterize their frequency-shift spectrum, simultaneous readout of all flux-tunable elements enables crosstalk characterization and compensation~\cite{Barrett2023,Dai_2021}. 

Meanwhile, flux-tunable qubits or couplers under flux crosstalk-induced frequency shifts may cause dispersive or ZZ-interaction shift with adjacent qubits, depending on the coupling strength. Adjacent qubits undergoing 1Q gate operations may experience frequency shifted, which affects their 1Q gate fidelity as discussed later.

\begin{figure}[htbp]
    \centering          \includegraphics[width=8.6cm]{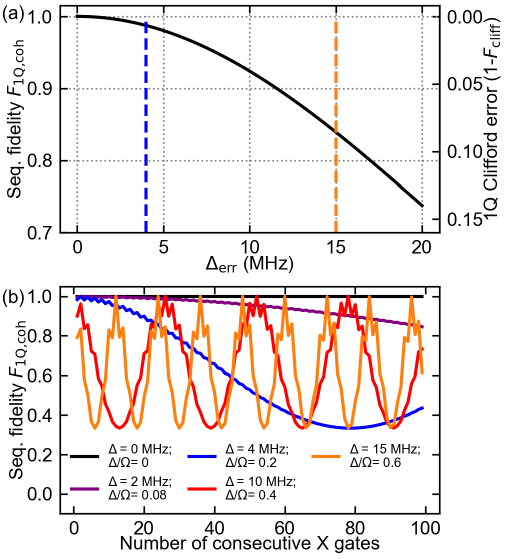}
    \caption{
    (a) Simulation of 1Q-gate fidelities, and single qubit gate Clifford errors as a function of detuning errors. Dashed line refers to the detunings observed for Q1 (Blue) and C1 (Orange) in Fig. 3(b) and 3(d) of the main text. The initial average state fidelities (average Clifford errors) for Q1 and C1 are 0.987 ($6.8\,\times\,10^{-3}$) and 0.83 ($8.6\,\times\,10^{-2}$), respectively. (b) Simulated dependences of 1Q-gate fidelities on the number of X-gate sequences at varying detuning errors.}
    \label{fig:s_sim1Qerr}
\end{figure}

In a connected multiqubit system, 1Q gates are affected when the probe frequency detunes from its nominally resonant XY driving frequency due to flux crosstalk caused by simultaneous biasing of flux lines. A two-level driven Hamiltonian can be used to simulate the effect of detuning $\Delta=\Delta_\text{err}$ on the average state fidelity of the qubits~\cite{Nielsen_2002,Tripathi_2025}: 
\begin{equation}
H=\frac{\Omega}{2}\sigma_x + \frac{\Delta_\text{err}}{2}\sigma_z,
\label{eqn:s_Hatom}
\end{equation}
where $\Omega$ is the Rabi frequency in angular units, and $\sigma_x$ and $\sigma_z$ are the Pauli matrices in the X and Z bases, respectively. The Hamiltonian evolves in time $t$ through the unitary operator
\begin{equation}
U=\text{exp}(-iHt).
\label{eqn:s_Unitary}
\end{equation}
Unitary operators allow us to study the control (coherent) errors without including the effect of decoherence. These operators help to determine the average 1Q gate fidelity~\cite{Nielsen_2002}
\begin{equation}
F_{1\text{Q,coh}}=\frac{[\text{Tr}(U^{\dagger} U_\text{ideal})]^2+d^2}{d^2(d + 1)},
\label{eqn:s_avgF1Q}
\end{equation}
where $d=2$ indicates the qubit states. To convert the average 1Q fidelity to Clifford errors $1-F_\text{cliff}$, we use the relation between 1Q gate errors and Clifford errors~\cite{Barends2014,Huang_2019}. 

The numerical simulation concentrates on X-gate fidelities, as demonstrated in previous works~\cite{Barrett2023}, including expansions to larger detunings. While Rabi drive amplitude errors cause rotation errors that shift the rotation angle of the X-gate, detuning errors lead to phase errors on the Bloch sphere, rotating the Bloch vector within the XY-plane. When the detuning has an equivalent phase error $\Delta_\text{err}/\Omega<1$\textdegree, the Clifford error rate per gate is minimal and undetectable by standard benchmarking protocols~\cite{Tripathi_2025}. At \SI{1}{\mega Hz} detuning, the fidelity decreases exponentially with increasing detuning~\cite{Barrett2023}. At significant large detuning, as seen in \fref{fig:s_sim1Qerr}(a), phase errors become significant enough to decrease both the average state fidelity and average Clifford error per gate at \SI{4}{\mega Hz} and \SI{15}{\mega Hz} detunings, respectively. The shape resembles the analytic fidelity for XX gates~\cite{Tripathi_2025} which is initially unity at small error limit and decays parabolically as phase errors increase. 

When the detuning approaches the Rabi frequency of $\Omega/2\pi=\,$\SI{25}{\mega Hz} under a square pulse duration $t_\text{XY}=20\,$ns, the resulting phase error tilts the Bloch vector into the XZ-plane~\cite{Barrett2023} and accumulate phase on the Bloch vector~\cite{Tripathi_2025}. As illustrated in \fref{fig:s_sim1Qerr}(b), the phase errors lead to oscillations in the average fidelity as the number of X-gate increases. This oscillatory pattern, found elsewhere~\cite{Tripathi_2025}, becomes more rapid with larger detunings. The amplitude is not necessarily maximal at the initial X-gate, since the Bloch vector projection gradually shifts toward the Z-axis in a conical trajectory. 

Phase error causes memory-like effects that accumulate over successive quantum gate operations~\cite{Tripathi_2025}.  Coherent errors are reversible using error-mitigation strategies, whereas decoherence and dephasing cannot. However, imperfect calibration of these errors can be detrimental in subsequent quantum circuits as they may induce destructive and constructive interference in gate compilations~\cite{Tripathi_2025}, as highlighted in \ref{fig:s_sim1Qerr}(b). Proper flux crosstalk calibration with qubits and couplers not only eliminates flux crosstalk-induced shifts, but also suppresses second-order effects like undesirable dispersive shifts and ZZ-interaction-based shifts~\cite{Barrett2023}. 

\section{\label{sec:sup_scale}Scalability of Flux Crosstalk Characterization and Compensation for Large-Scale Superconducting Processors} 

For MZLC to be adapted for characterizing large-scale superconducting processors, it should scale with the number of flux-tunable qubit and couplers in the system, while allowing to perform the calibration within reasonable amount of time. Here, we discuss several challenges in determining and inverting larger crosstalk matrices, as well as strategies to make the protocol more scalable.

\subsection{\label{sec:sup_scale_variation} Variations of Flux Crosstalk with Pulse Duration}

The discussion in Sec.~\ref{sec:sup_xtalk_vs_time} on the variations of the flux crosstalk with pulse duration sheds light on how flux-tunable qubits and couplers experience from transmon drifts due to long flux-settling time. Notably, flux-tunable transmons exhibit frequency shifts for short- and long-time flux transients below \SI{100}{ns}~\cite{TMLi2024,RLi2024}. Correcting both short- and long-time flux transients for all flux-sensitive qubits and couplers~\cite{Rol_2020,RLi2024,zhang2025_aswap} prior to flux crosstalk characterization is necessary; it stabilizes the total flux crosstalk with respect to flux pulse duration and enhances the robustness of the crosstalk matrix inversion. 

Resolving flux transient times for short- and long-term timescales depends on the cryoscope SNR, which is influenced by the readout fidelity, the flux-dependent qubit dephasing time $T_{2}^*(\Phi)$, the rate of the acquired phase, filtering effects in data processing, and the number of experimental runs used for averaging~\cite{Rol_2020,hellings_2025}. Cryoscope measurements typically take around a minute to capture dynamics up to \SI{200}{\nano s}. Meanwhile, our measurement in \fref{fig:s_xtalk_vs_time} took around a lower bound time of $1.42\,$h, which could have sped up using one-dimensional spectroscopic scans beyond \SI{100}{\nano s}~\cite{hellings_2025} per qubit. The overlap between cryoscope and spectroscopic techniques (as modeled in \eref{eqn:s_snr6}) relies on the number of averages and readout SNR. Higher readout SNR and good state preparation not only improve the resolution of frequency shifts caused by flux transients but also reduce the number of added experimental runs for averaging.

\subsection{\label{sec:sup_scale_char_time} Flux Characterization Time}

\begin{table}[htbp]
\caption{
Components of the Pulse Schedule used to characterize the 2D flux crosstalk map as shown in Figs. 3(b) to 3(g) and \ref{sfig:ramsey}(b). The FPGA time is $t_\text{FPGA}=\,$\SI{10}{\nano s}}
\begin{ruledtabular}
\begin{tabular}{cccccccc}
Method&
Probe& 
\makecell{$t_\text{init}\,$ \\ ($\mu$s)}&
\makecell{$t_\text{XY}\,$ \\ (ns)}&
\makecell{$t_{\Phi,\text{AC}}\,$ \\ (ns)}&
\makecell{$t_\text{buffer}\,$ \\ (ns)}&
\makecell{$t_\text{read}\,$ \\ ($\mu$s)}&
\makecell{$t_\text{sched}\,$ \\ ($\mu$s)}\\
\colrule
MZLC & Q & 200 & 0 & 140 & 20 & 0.4 & 200.57\\
MZLC & C & 200 & 0 & 140 & 20 & 4 & 204.17\\
Ramsey & Q & 200 & 40 & 100 & 60 & 0.4 & 200.61 \\
Ramsey & C & 200 & 40 & 100 & 60 & 4 & 204.21 \\
\end{tabular}
\end{ruledtabular}
\label{tab:xtalk_sched}
\end{table}

In scaling up flux crosstalk measurements, every duration in the pulse schedule including the XY, flux, and readout (RO) pulses, and buffer time $t_\text{buffer}$ plays an important role. The total pulse schedule time $t_\text{sched}$ includes not only the pulse schedule shown in Fig. \textcolor{blue}{3}(a) and \fref{sfig:ramsey}(b) but also the initialization time $t_\text{init}$ (also known as qubit reset time). The pulse schedule time reads:
\begin{equation}
t_\text{sched}= t_\text{init} + t_{\Phi_\text{AC}} + t_\text{buffer} + t_\text{read}+t_\text{FPGA}.
\label{eqn:s_tsched}
\end{equation}
where $t_\text{FPGA}=\,$\SI{10}{\nano s} is the buffer time allocated for FPGA bring-up after each pulse sequence. A detailed breakdown of the pulse schedule for both MZLC and the Ramsey methods is listed in Table \ref{tab:xtalk_sched}. The long initialization time and readout time dominate in both pulse schedules for qubits and couplers. Our current measurement uses passive reset for a $T_1=\,$\SI{20}{\micro s}, which sets $t_\text{init}$ to about $\text{max}\left(2\pi\,\times\,10\kappa^{-1}, 10\,\times\,T_{1}\right)$ to ensure that 1) the residual photons in the resonator are depleted, and 2) the qubit rests to ground state before the next sequence. The readout time for the coupler is significant longer than that of qubits because of the much weaker transmon-readout resonator coupling $g$ as summarized in Table \ref{tab:dev_params}. 

\begin{table}[htbp]
\caption{
Parameters used to characterize flux crosstalk using different
techniques, voltage range and resolutions, pixel sizes, averages and flux crosstalk characterization time}
\begin{ruledtabular}
\begin{tabular}{cccccccc}
Method&
Probe& 
\makecell{RO \\time\\(\text{$\mu$}s)}&
\makecell{S-Flux \\Res.\\ (mV)}&
\makecell{S-Flux \\Range \\ (V)}&
\makecell{Pixel \\ Count \\($n_\text{px}$)}& 
\textrm{$n_\text{avg}$}&
\makecell{$t_\text{xtalk}$\\(s)}\\ 
\colrule
MZLC & Q1 & 0.4 & 8 & 0.2 & 2500 & 100 & 51\\
MZLC & Q1 & 0.4 & 20 & 0.1 & 100 & 300 & 6\\
Ramsey & Q1 & 0.4 & 4 & 0.2 & 10000 & 100 & 210\\ 
MZLC & C1 & 4.0 & 8 & 0.2 & 2500 & 1000 & 513\\ 
MZLC & C1 & 4.0 & 20 & 0.2 & 400 & 400 & 35\\ 
Ramsey & C1 & 4.0 & 8 & 0.4 & 10000 & 100 & 213\\
\end{tabular}
\end{ruledtabular}
\label{tab:xtalk_params}
\end{table}

Table \ref{tab:xtalk_params} summarizes the measurement time required to characterize flux crosstalk for a given S-Flux range and resolution, as well as the corresponding number of pixels, and number of averages. All combinations lead to a benchmarked flux crosstalk characterization time. The total flux crosstalk characterization time $t_\text{xtalk}$ is given by    
\begin{equation}
t_\text{xtalk} = n_\text{avg} \times n_\text{px} \times t_\text{sched},
\label{eqn:st_xtalk}
\end{equation}
where $n_\text{px}=(2\,\times\,\text{range}/\text{resolution})^2$
and $n_\text{avg}\approx t_\text{xtalk}/(n_\text{px}\,\times\,t_\text{init})$. The pixel count of $n_\text{px}$ assumes an equal number of pixels for S-Flux and P-Flux axes. 

The MZLC scheme has a similar pulse schedule and measurement time to the Ramsey-based method, which are severely limited by $t_\text{init}$, as summarized in Tables~\ref{tab:xtalk_sched} and~\ref{tab:xtalk_params} respectively. The advantage of the MZLC method lies not in the characterization time but in its reduced overhead and robustness, which consequently leads to operational efficiency during processor bring-up. MZLC operates reliably using only coarsely calibrated pulse parameters to induce a measurable population change. This eliminates the mandatory, iterative overhead of precise $\pi$ and $\pi/2$ microwave pulse calibration, which are prerequisites for Ramsey-based measurements with high contrast. This robustness eventually decouples the flux crosstalk characterization node from the microwave pulse optimization node in the calibration stack. This decoupling allows the flux crosstalk matrix to be characterized and corrected immediately upon system bring-up, regardless of qubit-frequency fluctuations. Moreover, reducing the precise state preparation overhead in flux crosstalk calibration scales with the number of flux-tunable elements, enabling a more modular, robust, and scalable calibration workflow. Finally, we emphasize that the total characterization times reported in Table S4 are comparable for both protocols because they are primarily governed by identical systemic factors—including number of averages, initialization time, and substantial instrument and data processing latency—rather than the specific choice of measurement sequence.

\color{black}

Table~\ref{tab:xtalk_sched} allows us to estimate the lower bound of the total time needed to characterize the flux crosstalk matrix $t_\text{xmat}$ of our 2Q-2C processor. This estimate assumes serial, automated characterization of all off-diagonal elements, and the PPU characterizing the processor in a single fixed cabling session, and a time duration for linear regression of \SI{0.1}{\milli s}. For 12 off-diagonal elements, MZLC takes around $4.1-56.4$ minutes for low-resolution and high-resolution scans, respectively. 

Projecting this to a 100Q--180C processor with the current configuration would require the serial flux crosstalk matrix characterization time of
\begin{equation}
t_\text{xmat}=n_\text{elem}\times(n_\text{elem}-1)\times t_\text{xtalk}
\label{eqn:s_xmat_serial}
\end{equation}
where $n_\text{elem}$ is the number of flux-tunable and flux-sensitive elements in the processor. The total calibration time is therefore estimated to be approximately $133-1130\,$h. To make the protocol scalable would require lowering $t_\text{init}$. Reducing the criteria for a passive reset time to half ($5\,\times\,T_{1}$), while adapting to the maximum $T_{1}\approx\,$\SI{11}{\micro s} of the measured qubits, cuts $t_\text{init}=\,$\SI{55}{\micro s}, which cuts down $t_\text{xtalk}$ by a factor of 3.5. 

Another route for improvement is to enhance $\kappa$ either through Purcell Filter frequency targeting over the resonator~\cite{Sank2025}, or by using a flux-tunable Purcell filter~\cite{Xiong_2025}. This will allow one to reduce the readout time. Furthermore, a quantum-limited parametric amplifier is essential for further minimizing the added noise from the readout chain $n_\text{add}$. Following the notation in Ref.~\cite{Sank2025}, and with the measurement efficiency expressed in units of $n_\text{add}$~\cite{Gusenkova_2021}, we can write $\sqrt{\text{SNR}_\text{read}}$ for a transmon-resonator system with dispersive shift $\chi$, and number of average circulating photons $\bar n$ as:
\begin{equation}
\sqrt{\text{SNR}_\text{read}}=\sqrt{\frac{\bar n \kappa t_\text{read}}{n_\text{add}/2}}\left| \frac{4\chi/\kappa}{1+(2\chi/\kappa)^2} \right|.
\label{eqn:s_snr_model}
\end{equation}

For scalability, we set $\chi/\kappa=0.5$. The number of photons in a dispersive measurement is limited due to measurement-induced state transitions~\cite{Sank_2014,Sank2025}. Hence, $\kappa$ and $n_\text{add}$ can be improved without changing the transmon layout. Suppose the measurement efficiency of our current readout is 0.05~\cite{Bernier_2019}, corresponding to $n_\text{add}=19.5$~\cite{Bernier_2019,Gusenkova_2021}. By using typical values of $\kappa/2\pi=$\SI{16}{\mega Hz} with a well-targeted Purcell filter~\cite{Xiong_2025} while maintaining the ratio of $\chi/\kappa$, and getting a $n_\text{add}=8.7$ for a parametric amplifier with $20\,$dB gain~\cite{White_2023}, the measurement improves $\sqrt{\text{SNR}_\text{read}}$ by a factor of 18.1. If the target $\text{SNR}_\text{avg}$ remains the same, the improvement $\sqrt{\text{SNR}_\text{read}}$ reduces $\sqrt{n_\text{avg}}$ by a factor of 18.1, leading to time reduction by a factor of 330.

Introducing an active reset pulse, such as the CLEAR pulse~\cite{McClure_2016,Chatterjee_2025} reduces $t_\text{init}$ to less than \SI{1}{\micro s} for qubits with lifetime of hundreds of microseconds and $\chi/\kappa=0.5$. Moreover, reading out the coupler state through the hybridized mode between qubit and coupler~\cite{zhang2025_aswap} enables the coupler to share the same $t_\text{read}$ as the qubit. The combination of an optimized reset pulse and removal of buffer time reduces $t_\text{xtalk}=\,$\SI{1.31}{\micro s}, a factor of 153$\times$ reduction from the original $t_\text{xtalk}$ time.

Overall, introducing an optimized Purcell filter with precise frequency targeting, a $20\,$dB quantum-limited amplifier, and reshaped readout pulses for intrinsic reset reduces $t_\text{xtalk}$ by roughly $5.07\times10^{4}$. Consequently, the lower bounds of low resolution and high resolution scans of a 100$\,$Q-180$\,$C processor requires characterization times of $t_\text{xmat}=(2.6-22)\times10^{-3}\,$ hours. (or 9.5--80 s.) using serial characterization, which are reasonable time scales for a node in the calibration stack for quantum error correction stack~\cite{Mohseni_2025}.

\subsection{\label{sec:sup_scale_connect} Extension to Qubits and Tunable Elements with Larger Connectivities}

\begin{table}[htbp]
\caption{ 
Scaling of the number of flux-tunable elements involved in the flux crosstalk matrix is shown as a function of the number of elements. FTQ and FTC refer to flux-tunable qubits and couplers, respectively. FFQ refers to fixed-frequency qubits. $v$ refers to connectivity of the couplers. 2D Lattice refers to a square lattice, and parenthesis refers to evaluated qubit counts to 100. $n_\text{Q,2D}$ for the 2D lattice is defined as $n_\text{Q}=n_\text{Q,2D}^2$.}
\begin{ruledtabular}
\begin{tabular}{ccccc}
System &
\makecell{No. of FTQ \\  $n_\text{Q}$ \\ ($n_\text{Q}=100$)}& 
\makecell{No. of FTC \\ ($v=2$) \\ ($n_\text{Q}=100$)}&
\makecell{No. of FTC \\ ($v=4$) \\ ($n_\text{Q}=100$)}&
\makecell{$n_\text{elem}$}\\ 
\colrule
\makecell{1D Chain, \\ 
FTQ+FTC \\ $v=2$} & $n_\text{Q}$ (100) & $n_\text{Q}-1$ (99) & - & \makecell{$(2n_\text{Q}-1)^2$ \\ ($199^2$)} \\
\colrule
\makecell{1D Chain, \\ 
FFQ+FTC \\ $v=2$} & - & $n_\text{Q}-1$ (99) & - &  \makecell{$(n_\text{Q}-1)^2$ \\ ($99^2$)} \\
\colrule
\makecell{2D Lattice, \\ 
FTQ+FTC \\ $v=2$} & \makecell{$n_\text{Q,2D}^2$ \\ (100)} & \makecell{2$n_\text{Q,2D} \times $\\ $(n_\text{Q,2D}-1)$ \\ (180)}& - &  \makecell{$(3n_\text{Q,2D}^2$ \\$-2n_\text{Q,2D})^2$ \\ ($280^2$)} \\
\colrule
\makecell{2D Lattice, \\ 
FFQ+FTC \\ $v=2$} & - & \makecell{2$n_\text{Q,2D} \times $\\ $(n_\text{Q,2D}-1)$ \\ (180)}& - &  \makecell{$(2n_\text{Q,2D}^2$ \\$-2n_\text{Q,2D})^2$ \\ ($180^2$)} \\
\colrule
\makecell{2D Lattice, \\ 
FTQ+FTC \\ $v=4$} & \makecell{$n_\text{Q,2D}^2$ \\ (100)} & -& \makecell{$(n_\text{Q,2D}-1)^2$ \\ (81)} &  \makecell{$(2n_\text{Q,2D}^2$ \\$-2n_\text{Q,2D}$ \\ $+1 )^2$ \\ ($181^2$)} \\
\colrule
\makecell{2D Lattice, \\ 
FFQ+FTC \\ $v=4$} & - & - & \makecell{$(n_\text{Q,2D}-1)^2$ \\ (81)} &  \makecell{$(n_\text{Q,2D}$\\$-1)^4$ \\ ($81^2$)} 
\end{tabular}
\end{ruledtabular}
\label{tab:xtalk_scaling_elem}
\end{table}

So far, the discussion on flux-tunable qubits and couplers has focused on a qubit with connectivity of four and coupler connectivity of two. When extending our system to qubits and couplers with higher connectivities, we avoid discussion of the effect of complex wiring on the connectivities for simplicity. Any elements that are sensitive to flux bias and can act as crosstalk sources are considered part of the $n_\text{elem}\times n_\text{elem}$ flux crosstalk matrix. The total number of off-diagonal matrix elements is based on \eref{eqn:s_xmat_serial}:
\begin{equation}
n_\text{off-diag}=n_\text{elem}\times (n_\text{elem}-1)
\end{equation}

Table \ref{tab:xtalk_scaling_elem} lists combinations of qubits and couplers for specific systems: a 1D chain that restricts the connectivity of qubits to two, and a 2D lattice that expands the connectivity of qubits to four and couplers from two to four. When changing the number of flux-tunable qubits (FTQ) from a 1D chain to a 2D lattice, the number of flux crosstalk matrix does not increase. Hence, we expect flux-tunable qubits with higher connectivities, like in the case of Ref.~\cite{Renger_2025,Vigneau_2025}, the size of the qubit component of the flux crosstalk matrix remains unchanged.

However, for a flux-tunable coupler (FTC) case as shown in Table~\ref{tab:xtalk_scaling_elem}, the number of coupler elements increases significantly if the connectivity remains at $v=2$. The increasing FTC counts imply that for a 2D lattice, the number of coupler elements with $v=2$ dominates the flux crosstalk matrix compared to the qubits. Hence, a high-fidelity coupler state readout becomes very crucial for accurately assessing crosstalk effects in large processors. Increasing the coupler connectivity to $v=4$ as proposed in Ref~\cite{Miyazaki_2025}, reduces the number of flux crosstalk elements contributed by the coupler by more than half, potentially simplifying flux crosstalk characterization. Moreover, $n_\text{elem}$ drastically decreases when FFQs are used instead of FTQs.  

\subsection{\label{sec:sup_scale_spatial_var} Spatial Variations and Flux Crosstalk Level Threshold}

In constructing a flux crosstalk matrix for hundreds of qubits and couplers, understanding the spatial variation of the flux crosstalk $X_{ik}$ with respect to the Euclidean distance $r$ is essential for reducing the number of elements that require matrix inversion. The spatial dependence of the flux crosstalk is expressed as~\cite{Barrett2023}
\begin{equation}
|X_{ik}|(r)=\frac{100}{br+1}+h,
\label{eqn:sxtalk_r}
\end{equation}
where $b$ is an empirical constant that describes the reduction of flux crosstalk with increasing Eucledian distance from the diagonal elements $r$ and $h$ is the baseline crosstalk level that \eref{eqn:sxtalk_r} asymptotically approaches as $r=\infty$. \eref{eqn:sxtalk_r} is modeled so that $|X_{ik}|(r=0)=100$ corresponds to the 100$\,\%$ flux crosstalk from the diagonal elements. 

By curve-fitting a slice of one row of the flux crosstalk matrix, preferably one from the middle elements, using \eref{eqn:sxtalk_r}, one can judge whether $h+2\sigma_{h}$, where $2\sigma_{h}$ is twice the standard error of $h$ based on the fit, is approximately $-60\,$dB, corresponding to $0.1\,\%$ crosstalk error~\cite{Abdurakhimov_2024,Ihssen_2025}. If $h+2\sigma_{h}$ is above $0.1\,\%$ crosstalk error level or the saturated Euclidean distance $r_\text{sat}>n_\text{elem}/2$, then the entire flux crosstalk matrix must be characterized and compensated to achieve the flux crosstalk levels shown in Fig. 4(c) of the main text.

For specific device layouts and wirings like those in Refs.~\cite{Neill2018,Xiang_2023,Shi_2023_cal,Shi_2023_top}, the flux crosstalk matrices show block-diagonal patterns for flux crosstalk errors $|X_{ik}|<0.05$. These matrices can be block-diagonalized if the magnitudes across different blocks do not vary significantly, thereby greatly reducing the number of matrix elements that need to be calibrated for flux crosstalk.

Ref.~\cite{Barrett2023} further discussed a similar pulse technique (restricted in 1D scans with simultaneous readout), where a learning-based protocol was used to compensate flux crosstalk below $10\,\%$. Their simulations demonstrate that the median frequency error after flux crosstalk compensation increases when the nominal nearest-neighbor flux crosstalk exceeds $10\,\%$ based on 100 training sets. Beyond $10\,\%$, the flux crosstalk becomes nonlinear, and the linear flux crosstalk model used to describe and compensate off-diagonal flux crosstalk may not hold anymore. The $10\,\%$ could be the limiting crosstalk level for which MZLC remains valid. While the results in~Ref.~\cite{Barrett2023} may not directly apply to our measurement, investigation of the robustness and compensation limit of our technique may require more thorough studies. 

\bibliography{sup} 